\documentclass{article}

\usepackage{PRIMEarxiv}

\usepackage[utf8]{inputenc} 
\usepackage[T1]{fontenc}    
\usepackage{tikz}
\usepackage{algorithm}
\usepackage{amsfonts} 
\usepackage{algpseudocode}
\usepackage{multirow}
\usepackage{graphicx}
\usepackage{caption}
\usepackage{subcaption}
\usepackage{amsthm}
\usepackage{makecell}
\usepackage{url}
\usepackage{comment}
\usepackage{mathtools}
\usepackage{booktabs}
\usepackage{soul}
\DeclarePairedDelimiter\ceil{\lceil}{\rceil}

\newtheorem*{remark}{Remark}

\newtheorem{innercustomthm}{Theorem}
\newenvironment{mythm}[1]
  {\renewcommand\theinnercustomthm{#1}\innercustomthm}
  {\endinnercustomthm}
  
\newcommand\MyBox[2]{
  \fbox{\lower0.75cm
    \vbox to 1.7cm{\vfil
      \hbox to 1.7cm{\hfil\parbox{1.4cm}{#1\\#2}\hfil}
      \vfil}%
  }%
}   

\definecolor{mydarkblue}{HTML}{073b4c}

\usepackage{pgfplots}

\title{Certified Adversarial Robustness of Machine Learning-based Malware Detectors via (De)Randomized Smoothing
}

\author{
  Daniel Gibert \\
  CeADAR, University College Dublin \\
  Dublin, Ireland \\
  \And
  Luca Demetrio \\
  University of Genova \\
  Genova, Italy \\
   \And
  Giulio Zizzo \\
  IBM Research Europe \\
  Dublin, Ireland \\
   \And
  Quan Le \\
  CeADAR, University College Dublin \\
  Dublin, Ireland \\
   \And
  Jordi Planes \\
  University of Lleida \\
  Lleida, Spain \\
  \And
  Battista Biggio\\
  University of Cagliari\\
  Cagliari, Italy\\
}

\begin{document}
\maketitle

\begin{abstract}
Deep learning-based malware detection systems are vulnerable to adversarial EXEmples - carefully-crafted malicious programs that evade detection with minimal perturbation. 
As such, the community is dedicating effort to develop mechanisms to defend against adversarial EXEmples. However, current randomized smoothing-based defenses are still vulnerable to attacks that inject blocks of adversarial content.
In this paper, we introduce a certifiable defense against patch attacks that guarantees, for a given executable and an adversarial patch size, no adversarial EXEmple exist. Our method is inspired by (de)randomized smoothing which provides deterministic robustness certificates. During training, a base classifier is trained using subsets of continguous bytes. At inference time, our defense splits
the executable into non-overlapping chunks, classifies each chunk independently, and computes the final prediction through majority voting to minimize the influence of injected content. Furthermore, we introduce a preprocessing step that fixes the size of the sections and headers to a multiple of the chunk size. As a consequence, the injected content is confined to an integer number of chunks without tampering the other chunks containing the real bytes of the input examples, allowing us to extend our certified robustness guarantees to content insertion attacks. We perform an extensive ablation study, by comparing our defense with randomized smoothing-based defenses against a plethora of content manipulation attacks and neural network architectures. Results show that our method exhibits unmatched robustness against strong content-insertion attacks, outperforming randomized smoothing-based defenses in the literature.
\end{abstract}

\keywords{Malware Detection \and Adversarial Malware Examples \and Certified Defence \and Adversarial Learning \and (De)Randomized Smoothing}

\section{Introduction}

During the past decade, there has been a rise in various types of cyber threats, from malware to social engineering tactics like phishing attacks. According to the European Union Agency for Cybersecurity (ENISA)~\footnote{\url{https://www.enisa.europa.eu/publications/enisa-threat-landscape-2023}}, malware, and particularly ransomware, are among the prime cyber threats that face organizations today, with ransomware incidents surging in the first half of 2023 while showing no signs of slowing down. To keep up with malware and to overcome the prior pitfalls of traditional antivirus engines in detecting zero-day malware, researchers and cyber security companies started adopting machine learning to complement their solutions due to its suitability for processing large volumes of data~\cite{GIBERT2020102526}. Even though feature-based detectors~\cite{anderson2018ember,10.1145/2857705.2857713} are still prominent and are the most common type of malware detectors, there has been a notable shift towards deep learning-based or end-to-end detectors~\cite{DBLP:conf/aaai/RaffBSBCN18,DBLP:conf/iclr/KrcalSBJ18,GIBERT2021102159,DBLP:conf/aaai/RaffFZAFM21} in recent years. This shift has been primarily fueled by the unprecedented success of deep learning models in various domains, including computer vision, natural language processing, and now, cybersecurity. In addition, unlike feature-based detectors that rely on manually crafted features, deep learning-based detectors automatically learn relevant features and representations from raw bytes, simplifying the detection pipeline.

However, while promising, the use of deep learning to detect and classify malware presents attackers with novel opportunities to circumvent detection. Adversarial attacks, specifically designed to deceive deep learning models, have emerged as a significant concern. 
Subtly manipulating malware examples by injecting and optimizing adversarial content through techniques such as the Fast Gradient Sign Method~\cite{DBLP:conf/sp/SuciuCJ19,demetrio2021adversarial} and genetic algorithms~\cite{demetrio2021functionality,YUSTE2022102643}, attackers can create adversarial EXEmples that retain malicious intent but evade detection by fooling the target malware detectors.

As a result, the community has been dedicating substantial efforts, although with limited success, in developing defense mechanisms that mitigate the risks posed by adversarial EXEmples. Techniques like adversarial training have been shown ineffective in defending against SOTA attacks not seen during training~\cite{advtrain:sec2023}. Furthermore, various practical randomized smoothing-based defenses~\cite{gibert2023_randomizedsmoothing,huang2023rsdel} have been recently proposed to defend against adversarial EXEmples. However, neither defense is effective against SOTA attacks that inject localized adversarial payloads within a Portable Executable. Moreover, the probabilistic certified defense proposed by Huang et al.~\cite{huang2023rsdel} only achieves a certified accuracy of 91\% at an edit distance radius of 128 bytes, insufficient to cover the attack space of SOTA attacks which is usually in the order of thousands of bytes.

In this paper, we aim to address the limitations of existing defense mechanisms against adversarial EXEmples in malware detection. To this end, we introduce a novel smoothing-based classification scheme specifically designed to defend against content manipulation attacks. Our approach draws inspiration from (de)randomized smoothing~\cite{DBLP:conf/nips/0001F20a}, a certified defense against patch attacks originally designed for image classifiers. Our scheme works as follows: (1) At training time, a base classifier is trained using only a single subset of contiguous bytes or chunk of bytes from an executable; (2) At test time, our defense splits the executable into
non-overlapping chunks and classifies each chunk independently, and predictions are computed through majority voting to reduce the influence of the injected content. This allows us to derive a deterministic robustness certificate for our classifier under different types of attacks if the number of predictions for the correct class exceeds the second most commonly predicted class by a large enough margin. As far as we are aware, no other deterministic certified defense against content manipulation attacks on deep learning-based malware detectors has been proposed so far.

This work extends our previous work in~\cite{10.1145/3605764.3623914} by:
\begin{itemize}
    \item Extending the deterministic certified defense against functionality-preserving content manipulation attacks on deep learning-based malware detectors.
    \item Introducing a preprocessing step that allows us to provide deterministic robustness certificates against patch, append, and content injection attacks.
    \item Rigorously evaluating our certified defense across three neural network architectures, namely MalConv~\cite{DBLP:conf/aaai/RaffBSBCN18}, MalConvGCT~\cite{DBLP:conf/aaai/RaffFZAFM21}, and AvastConv~\cite{DBLP:conf/iclr/KrcalSBJ18}.
    \item Performing an extensive ablation study, comparing our certified defense with other state-of-the-art defenses against a plethora of content manipulation attacks, establishing a new benchmark in terms of adversarial and certified accuracy against adversarial EXEmples.
    \item Showing how the proposed smoothing scheme can be used to facilitate a finer-grained analysis of the file, by identifying specific chunks within a file that exhibit malicious traits.
\end{itemize}

The rest of the paper is organized as follows. Section~\ref{sec:related_work} provides a review of the related work that address the robustness of deep learning-based malware detectors against adversarial EXEmples. Section~\ref{sec:derandomized_smoothing} presents our deterministic certified defense. Section~\ref{sec:evaluation} evaluates the proposed defense mechanism against various state-of-the-art evasion attacks. Section~\ref{sec:discussion} discusses the main strengths and weaknesses of the proposed defense with respect to other state-of-the-art defenses. Finally, Section~\ref{sec:conclusions} summarizes our concluding remarks and presents some future lines of research.


\section{Related Work}
\label{sec:related_work}
Before delving into our methodology, we first review related works that address robustness against adversarial EXEmples.

Fleshman et al.~\cite{https://doi.org/10.48550/arxiv.1806.06108}  proposed non-negative weights to defend against benign content injection. The drawback of the non-negative model is that its accuracy slightly decreases compared to the standard model on the original executables due to its inability to use features that correspond to benign software. In addition, it has been proven that a non-negative network trained with softmax activation can be transformed into an equivalent constrained network~\cite{6783731} breaking this style of defence.\footnote{\,Unfortunately non-negative MalConv does not have public model weights or source code. In addition, the non-negative MalConv model provided for the Machine Learning Static Evasion Competition is faulty as it was trained using the softmax activation. We were unable to replicate the results of the non-negative MalConv paper. Hence, we decided to leave the non-negative model out of our experiments due to lack of reproducibility.}

Lucas et al.~\cite{advtrain:sec2023} proposed to augment the training data with weak versions of three evasions attacks to train a robust end-to-end classifier. In their work, the attacks considered are the following: (1) In-Place Replacement attack (IPR)~\cite{10.1145/3433210.3453086}, (2) Displacement attack (Disp)~\cite{10.1145/3433210.3453086} and (3) Padding attack~\cite{DBLP:conf/sp/SuciuCJ19}. However, adversarial training demands considerable computational resources and requires knowledge of the type of modifications applied by the attacker to generate the malware examples. In addition, adversarial training does not guarantee robustness against attacks that significantly differ from the ones used during training~\cite{DBLP:conf/iclr/ZhangCSBDH19}.  

Recently, various defenses~\cite{huang2023rsdel,gibert2023_randomizedsmoothing} have been proposed to defend against adversarial attacks based on randomized smoothing~\cite{DBLP:conf/icml/CohenRK19,DBLP:conf/nips/LiCWC19,DBLP:conf/sp/LecuyerAG0J19}. Randomized smoothing is a technique that increases the robustness of ML classifiers against adversarial attacks by adding random noise to the input data during both training and inference, with the goal of making the classifier's predictions stable and resilient to small perturbations. By doing this, the model becomes more resilient to adversarial attacks because the added noise helps to ``smooth out” small changes in the input data that an attacker might make. Unlike adversarial training, smoothing-based classifiers focus on blurring the decision boundaries within the input space, improving the classifier's resilience against a broader range of attacks~\cite{DBLP:conf/icml/CohenRK19}. This method has recently gained attention in the computer vision domain due to its capacity to reduce a model’s sensitivity to noise or fine-scale variations. More specifically, it has been shown that randomized smoothing schemes employing Gaussian noise when randomizing inputs, provide $l_{p}$ robustness certificates~\cite{DBLP:conf/icml/CohenRK19,DBLP:conf/nips/LiCWC19,DBLP:conf/sp/LecuyerAG0J19}

Randomized smoothing works as follows: at training time, a base classifier $f$ is trained on ablated versions $\tilde{x}$ of a given input example, where $\tilde{x}$ consists of a noisy version of the input example $x$. At inference time, $L$ smoothed versions of an input example $x$ are generated and the final classification is determined as the class most frequently predicted across the set of ablated versions of the input example. 

Gibert et al.~\cite{gibert2023_randomizedsmoothing} presented a smoothing mechanism based on random byte substitutions. In their work, the ablated examples are generated by randomly substituting the bytes from an input example $x$ with probability $p$ for a specially-encoded 'PAD' token. By doing so, the proposed defense method achieved higher robustness against adversarial malware examples compared to a baseline classifier at the expense of increased computational time.

Alternatively, Huang et al.~\cite{huang2023rsdel} proposed RS-Del, a smoothing mechanism based on random deletions. Their method involves randomizing the deletion of bytes and analyzing this affects the classifier's predictions to confer probabilistic robustness certificates against adversarial deletion, insertion and substitution (patch) edits. In their work, instead of ablating bytes given a probability $p$ they deleted bytes. However, the proposed deletion mechanism only confers meaningful probabilistic robustness certificates against small edit attacks, i.e. certified accuracy of 91\% at an edit distance radius of 128 bytes, and many attackers are able to use larger edit distance and through SOTA attacks such as the Slack~\cite{DBLP:conf/sp/SuciuCJ19}, the GAMMA~\cite{demetrio2021functionality} and the Disp attack~\cite{10.1145/3433210.3453086} can bypass detection.

\section{(De)Randomized Smoothing for Malware Detection}
\label{sec:derandomized_smoothing}

Randomized smoothing, while effective against certain types of attacks~\cite{DBLP:conf/icml/CohenRK19,DBLP:conf/nips/LiCWC19,DBLP:conf/sp/LecuyerAG0J19}, has been shown to be ineffective in defending against adversarial EXEmples generated with SOTA evasion attacks~\cite{DBLP:conf/sp/SuciuCJ19,demetrio2021functionality,10.1145/3433210.3453086,demetrio2021adversarial}. The distinctive constraints imposed by executable files present unique challenges, leading to differences in the nature of adversarial attacks within the malware domain compared to attacks in the image domain.

In the context of malware detection, the bytes within executable files serve functional purposes and attackers commonly introduce the adversarial payloads into specific locations of executable files, rather than making arbitrary alterations to bytes. Unlike the manipulation flexibility found in the image domain, where attackers can alter any pixel in an image at will, indiscriminate modifications to the bytes in a file can disrupt the intended functionality of the executable and potentially render it nonfunctional. 

Due to this distinct nature of adversarial attacks in the realm of malware, where adversarial payloads are injected in specific locations of files, the application of randomized smoothing, employing random noise, may not be the most suitable defense mechanism against adversarial EXEmples.

\begin{figure*}[ht]
    \includegraphics[width=0.8\textwidth]{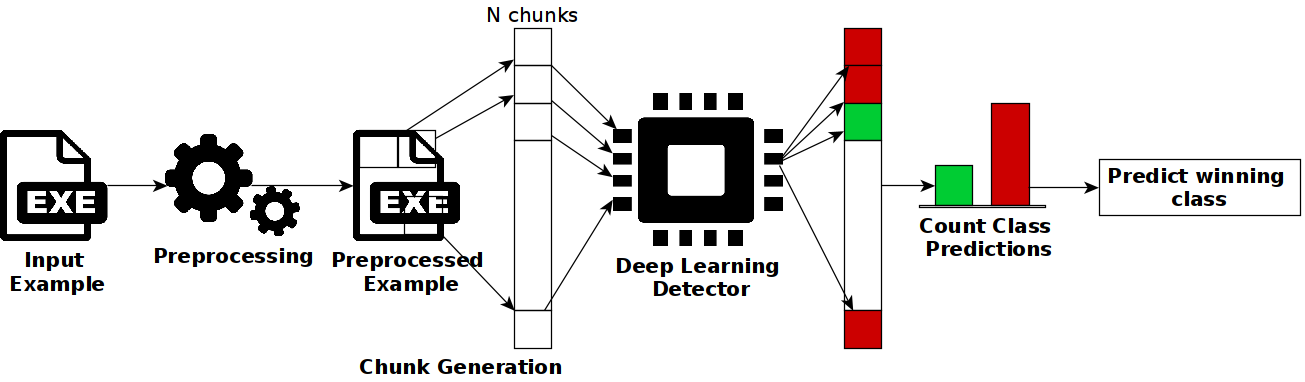}
    \centering
    \caption{An overview of the chunk-based smoothing classification scheme.}
    \label{fig:chunk_ablation_scheme_overview}
\end{figure*}

To overcome the limitations of randomized smoothing, we propose a smoothing-based defense strategy specifically designed against adversarial EXEmples. Our approach draws inspiration from (de)randomized smoothing~\cite{DBLP:conf/nips/0001F20a}, a class of certified robust image classifiers designed to defend against patch attacks, and our method adapts the concept for malware detection. The proposed defense strategy employs a chunk-based smoothing scheme, where instead of randomly replacing~\cite{gibert2023_randomizedsmoothing} or randomly removing bytes~\cite{huang2023rsdel} from a file, we ablate whole subsequences of bytes to create a more robust classifier.

In our chunk-based smoothing scheme, a base malware classifier, $f$, is trained to make classifications based on an ablated version $\tilde{x}$ of a given executable file $x$. This ablated version $\tilde{x}$ consists of only a subset of contiguous bytes or chunk of bytes of size $z$ randomly selected from the original file $x$. The rest of the file is is not used for classification. This functionality is implemented by the operation $\text{ABLATE}_{train}(x,z)$. The training procedure, defined in Algorithm~\ref{alg:smoothed_classifier_training}, involves training a base malware classifier with ablated versions of the files in the training dataset. Notice that during training, the starting locations of the chunks of bytes are randomly selected. For a description of the operation $\text{PREPROCESS}(x)$ we refer the readers to Section~\ref{sec:insertion_attacks}.
\begin{remark}
    In our work, instead of ablating the bytes with a NULL value like in Levine et al.~\cite{DBLP:conf/nips/0001F20a}, we simply remove the bytes. Removing the bytes ensures that no artificial placeholders are introduced and reduces the input size of the classifier to only that of the selected chunk, resulting in greater computational efficiency while maintaining similar detection performance.
\end{remark}


\begin{algorithm}
\caption{Smoothing-based classifier's training procedure.}\label{alg:smoothed_classifier_training}%
\begin{algorithmic}
\Require training dataset $D_{train}$, malware detector $f$ with parameters $\theta$, chunk size $z \in \mathbb{N}_{+} \: ,\:  z>0$
\State $\theta \gets Initialize\: parameters $
\For{i=1, MAX\_EPOCHS}
    \For{$x, y \in D_{train}$}
        \State $\hat{x} \gets \text{PREPROCESS}(x, z)$
        \State $\tilde{x} \gets \text{ABLATE}_{train}(\hat{x}, z)$
        \State $\tilde{y}   \gets f(\tilde{x}, \theta )$
        \State $Loss \gets criterion(y, \tilde{y} )$
        \State $\theta \gets Update\:  parameters$
    \EndFor
\EndFor
\end{algorithmic}
\end{algorithm}

At test time, an input example $x$ is split into consecutive, non-overlapping chunks of size $z$ starting from the byte located at position zero, i.e. $x=\{ \tilde{x_0}, \tilde{x_1}, \tilde{x_2}, ..., \tilde{x_N}\}$, where $\tilde{x_b}$ is a chunk of bytes from $x$ starting and ending at byte $b*z$ and $(b+1)*z$, respectively. Afterwards, each chunk $\tilde{x_b}$ is independently classified by the base classifier $f$. In other words, $f(\tilde{x_b})$ is the base classification, where the classifier takes as input the chunk of bytes $\tilde{x_b}$. For the task of malware detection, $C = \{c_1, c_2\}$ is the set of all possible classes, where $c_1$ and $c_2$ corresponds to the benign and malicious classes, respectively. For each class $c\in C$, $f_c(\tilde{x_b})$ will either be 0 or 1. To make the final classification and compute the robustness certificate, the number of chunks on which the base classifier returns for each class is counted:
\begin{equation}
    \forall_c \in C, n_c(x) \coloneqq \sum_{b=0}^{N} f_c (\tilde{x_b})
\end{equation} 

The final smoothed classification is simply the plurality class returned: 
\begin{equation}
    g(x) := \underset{c}{\text{argmax}}\: n_c(x)
\end{equation} In
the case of ties, we deterministically return the larger-indexed class, which corresponds to the malicious class. 

The probability $P(c|x)$ of an input example $x$ belonging to class $c$ can be estimated as:
\begin{equation}
    P(c|x) = \frac{n_c(x)}{N}
\end{equation}
where $n_c(x)$ is the number of chunks of $x$ that belong to class $c$ according to the base classifier $f$ and $N$ is the total number of chunks.

\begin{remark}
    Notice that our approach can be easily extended to a multi-class classification problem such as the task of classifying malware into families. In this case, the cardinality $|C|$ will be equal to the number of malware families. In the context of malware classification, if there is a tie, $f$ would return a predetermined class or the smaller-indexed class. In the context of malware family classification, it would not be necessary to require that $f_c(\tilde{x_b})=1$ for any class $c$ as the base classifier may abstain, returning $0$ for all classes when the probability of belonging to class $c$ is not above a predefined threshold, and could also return 1 for multiple classes. We would like to denote that this functionality hasn't been implemented as our approach focuses only on the task of malware detection. 
\end{remark}

\subsection{Certifiably Robust Classifier Against Content Manipulation Attacks}
We follow a similar reasoning to Levine et al.\,\cite{DBLP:conf/nips/0001F20a} to obtain a deterministic robustness certificate for our classifier under different types of attacks if the number of predictions for the correct class exceeds the second most commonly predicted class by a large enough margin.

Assuming an input example $x$ has size $L$ and a chunk size $z$, our smoothing strategy will split the classifier into $\ceil*{\frac{L}{z}}$ non-overlapping chunks of size $z$, except for the last chunk, which might be smaller and will be padded accordingly. Contrarily to randomized smoothing-based defenses~\cite{huang2023rsdel,gibert2023_randomizedsmoothing}, with our smoothing strategy it is tractable to use the base classifier $f$ to classify all possible ablated versions of an input example $x$, allowing us to exactly compute the smoothed classifier $g(x)$, yielding deterministic certificates. 
Leveraging the fact that an adversary has influence over a limited number of chunks, if the number of predictions of the correct class exceeds the other class predictions by a significant enough margin, we can certify the prediction of the smoothed classifier.

Following this reasoning, we can certify the prediction of our malware classifier against two types of attacks, (1) patch attacks, and (2) insertion attacks, where insertion attacks include padding attacks and injection attacks. Depending on the type of attack, the margin between the number of predictions for the correct class and the others will differ.

\subsubsection{Patch Attacks}
In the context of malware detection, a patch attack involves manipulating and replacing existing bytes within a given executable without inserting new bytes or removing existing bytes.
If the attacker perturbs a byte file with an adversarial patch of size $p > 1$, they can intersect with at most 
$\Delta = \ceil*{\frac{p}{z}}+1$ chunks, and thus, it can only alter the prediction of $\Delta$ chunks. This yields the following guarantee:
\begin{mythm}{1}
\label{theorem:certification_patch_attack}
    For any input example $x$, base classifier $f$, smoothing classifier $g$, smoothing chunk size $z$, and patch size $p$, such that $\Delta = \ceil*{\frac{p}{z}}+1$, if:
    \begin{equation}
       n_{c^{'}}(x) \ge  \underset{c^{''} \neq c^{'}}{max}  \left[n_{c^{''}}(x) +1_{c^{'}<c^{''}}\right]+ 2\Delta
    \end{equation}
    where $n_{c^{'}}(x)$ and $n_{c^{''}}(x)$ denote the most commonly and second most commonly predicted classes for an input example $x$, then for any adversarial EXEmple $x^{'}$ which differs from $x$ only in a patch of size $p$, $g(x^{'})=c^{'}$.
\end{mythm}
In Theorem~\ref{theorem:certification_patch_attack}, the indicator function term ($1_{c^{'}<c^{''}}$ ) is present because we break ties deterministically by label index during the final classification. Proofs are provided in Appendix~\ref{sec:proofs}.


Since the problem of malware detection is a binary classification problem, this can be rewritten as follows for predicting an input file as malware:
\begin{equation}
    n_{m}(x) \ge  n_{b}(x) + 2\Delta 
\end{equation}
where $n_{m}(x)$ and $n_{b}(x)$ refer to the number of chunks predicted as malware and benign by the smoothed classifier, respectively.

If instead of perturbing a single adversarial patch the attacker perturbs a byte file with $K$ adversarial patches of sizes $p_1, p_2, ..., p_K$, then $\Delta$ can be rewritten as follows:
\begin{equation}
\label{formula:certified_robustness_patch_attack}
    \Delta = \sum_{i=1}^{K} \Delta_i = \sum_{i=1}^{K} (\ceil*{\frac{p_i}{z}} +1)
\end{equation}
where $\Delta_i$ is the number of chunks the $i$-th adversarial patch can intersect with and $p_i$ is the size of the $i$-th adversarial patch.
Generally speaking, the decision of the classifier will only change depending on the amount of chunks that the attacker modifies. If the number of chunks that the attacker modifies is not large enough, then we can certify that the prediction of our classifier will not change. We summarize the certification procedure for a single example in Algorithm~\ref{alg:certification_procedure}. 

\begin{algorithm}
\caption{Certify if patch (or insertion) attack can change the predicted class of a malicious example $x$}\label{alg:certification_procedure}%
\begin{algorithmic}
\Require Example $x$, label $y$, chunk-based malware detector $f$
\For{$i=1, \dots , N$}
    \State set $\tilde{x_{i}}$ as a chunk of $x$ starting at position $z*i$ and ending at position $z*(i+1)$
    \State $\tilde{y_{i}} = f(\tilde{x_{i}})$ // Predicted class of chunk $\tilde{x_{i}}$ 
\EndFor
\State $\tilde{y}^{'}, \tilde{y}^{''} = $ majority and second majority of $\left \{ \tilde{y}_{i} \right \}_{i=1}^{n}$
\State $n_{c^{'}}(x), n_{c^{''}}(x) = $ number of chunks classified as $\tilde{y}^{'}, \tilde{y}^{''} $, respectively

\If{$ n_{c^{'}}(x) \ge  n_{c^{''}}(x) + 2\Delta $} // $ n_{c^{'}}(x) \ge  n_{c^{''}}(x) + \Delta $ for insertion attacks
    \State return certified
\Else
    \State return not certified
\EndIf

\end{algorithmic}
\end{algorithm}

\subsubsection{Insertion Attacks}
\label{sec:insertion_attacks}
Content insertion attacks refer to those attacks that insert new content within an executable. Insertion attacks can be divided into two: (1) append attacks and (2) injection attacks. On the one hand, append attacks, involve adding extra bytes or the adversarial payload at the end of a given executable, known as its overlay. This manipulation has long been used by malware authors to hide malicious code or data within seemingly benign files. On the other hand, injection attacks insert the adversarial payload somewhere within the executable file. This can be done by injecting the adversarial payload into newly-created sections~\cite{demetrio2021functionality}, creating space between the headers and the sections~\cite{demetrio2021adversarial}, or extending the space between sections~\cite{YUSTE2022102643}.



Similarly to patch attacks, if the attacker perturbs the file by appending an adversarial payload of size $p$, the maximum number of chunks from the adversarial EXEmple 
$x^{'}$ that will differ from $x$ will be $\Delta = \ceil*{\frac{p}{z}}+1$ chunks. This yields the following guarantee:
\begin{mythm}{2}
    \label{theorem:certification_append_attack}
    For any input example $x$, base classifier $f$, smoothing classifier $g$, smoothing chunk size $z$ and adversarial payload of size $p$, such as $\Delta = \ceil*{\frac{p}{z}}+1$, if:
    \begin{equation}
       n_{c^{'}}(x) \ge \underset{c^{''} \neq c^{'}}{max} \left[n_{c^{''}}(x) +1_{c^{'}<c^{''}}\right]+ \Delta
    \end{equation}
    where $n_{c^{'}}(x)$ and $n_{c^{''}}(x)$ denote the most commonly and second most commonly predicted classes for an input example $x$, then for any adversarial EXEmple $x^{'}$ which differs from $x$ only in an adversarial payload of size $p$, $g(x^{'})=c^{'}$.
\end{mythm}

This can be rewritten as follows:
\begin{equation}
\label{formula:certified_robustness_append_attack}
    n_{m}(x) \ge  n_{b}(x) + \Delta
\end{equation}

Notice that for append attacks the margin between the correct class and the others does not require to be as large as in patch attacks. This is because append attacks modify the executable by appending code at the end, and thus, the predictions for the chunks consisting of the original code of the executable won't be modified, except one chunk. The only case where no predictions for existing chunks are affected is when the chunks of the original example line up exactly to a multiple of the chunk size, otherwise the final chunk prediction is affected.  

\begin{figure*}[ht]
   \includegraphics[width=0.9\textwidth]{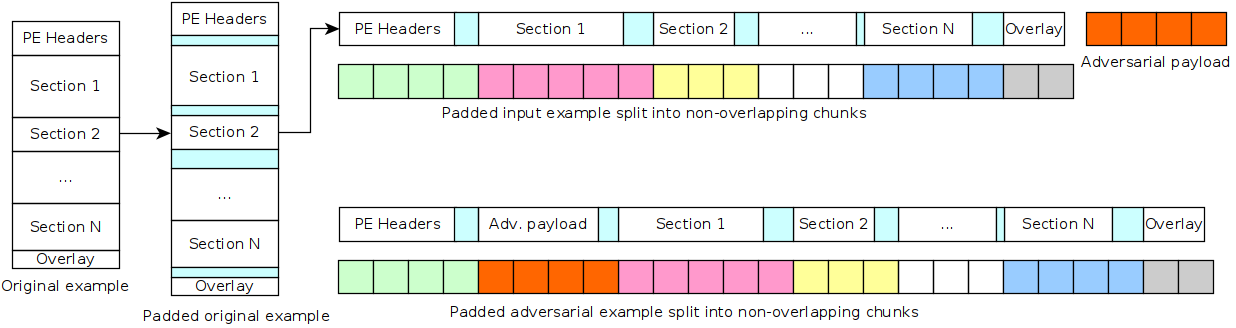}
    \centering
    \caption{An illustration of the $\text{PREPROCESS}$ operation. The $\text{PREPROCESS}$ operation pads the headers and sections of Portable Executable files to a multiple of the chunk size $z$ used to split the executable into non-overlapping chunks. By doing so, we ensure that the predictions for the chunks corresponding to the headers and sections not manipulated by the attacker in both the original and adversarial EXEmples remain consistent, allowing us the extend the certification procedure to content injection attacks.}
    \label{fig:preprocess_functionality}
\end{figure*}

Unfortunately, the certification procedure employed for append attacks does not directly translate to injection attacks. The reason behind this disparity is that injection attacks introduce new content within the executable, and not at the end, disrupting the file's original structure. Consequently, the content within the original file gets displaced, leading to discrepancies in the content of the chunks from the original and the adversarial EXEmples as seen in Figure~\ref{fig:preprocess_functionality}. 

To address this challenge, we have implemented a preprocessing step, denoted as $\text{PREPROCESS}$. This step takes a given example as input and appends zero bytes strategically to the headers and sections in a manner that aligns them with multiples of the chunk size used for splitting the executable. By doing so, we ensure that the predictions for the chunks corresponding to each section in both the original and adversarial EXEmples remain consistent. This alignment enables us to extend our certification approach, originally designed for append attacks, to assess the robustness of our system against injection attacks using Theorem~\ref{theorem:certification_append_attack} or Equation~\ref{formula:certified_robustness_append_attack}.
Furthermore, if the attacker inserts more than one adversarial payload with sizes $p_1, p_2, ...,p_k$, then $\Delta$ can be rewritten as using Equation~\ref{formula:certified_robustness_patch_attack}.

\section{Evaluation}
\label{sec:evaluation}
We now detail the setup of our experiments, by specifying the hardware we used to evaluate our methodology, the datasets we have considered, and the baselines models and neural networks that we train with out certification schema.
\subsection{Experimental Setup}
The experiments have been run on a machine with an Intel Core i7-7700k CPU, 1xGeforce GTX1080Ti GPU and 64 GB RAM. The code has been implemented with PyTorch~\cite{NEURIPS2019_9015} and will be publicly available in our Github repository after acceptance~\footnote{\url{https://github.com/danielgibert/certified_robustness_of_end_to_end_malware_detectors}}.
\subsubsection{BODMAS dataset}
In this paper, we have assessed the proposed chunk-based smoothing scheme using the BODMAS dataset~\cite{bodmas}. This dataset comprises a total of 57,293 malware samples, including information on 581 malware families, along with 77,142 benign Windows PE files collected between August 2019 and September 2020. The dataset has been partitioned into three sets, training, validation and test sets, based on the timestamps associated with each sample. As a consequence, the training set contains older executables, while the test set comprises the most recent executables.

To speed-up our experiments, we have limited our focus to executables that are either equal to or smaller than 1 megabyte. This choice is driven by the computational demands associated with larger input sizes, as well as the need to manipulate executables by injecting and optimizing with genetic algorithms~\cite{demetrio2021functionality,YUSTE2022102643}. By restricting our analysis to executables of 1 megabyte or less, we eliminate the necessity to trim bigger executables, ensuring that no critical information is lost during the classification, while streamlining the training and testing process, ensuring efficient and timely model development and evaluation. Consequently, the resulting reduced dataset consists of 39,360 benign and 37,742 malicious executables~\footnote{The SHA-256 hashes associated with our subset of examples from the BODMAS dataset will be provided to allow researchers to reproduce our work}.

\subsubsection{Deep Learning-based Malware Detectors}
In the context of malware detection, convolutional neural networks have been typically applied to build end-to-end models that learn to classify executable files directly from their raw byte sequences.
In this work, we experiment with the following convolutional neural network architectures.
\begin{itemize}
    \item \textit{MalConv}~\cite{DBLP:conf/aaai/RaffBSBCN18}: this architecture 
    consists of an embedding layer, a gated convolutional layer, a global-max pooling layer and a fully-connected layer.
    \item \textit{MalConvGCT}~\cite{DBLP:conf/aaai/RaffFZAFM21}: this architecture 
    consists of two sub-networks, a context extractor and a feature extractor sub-network. On the one hand, the context extractor sub-network learns global context from the input examples. On the other hand, the feature extractor sub-network performs initial feature extraction. Following, a global channel gating layer is used to selectively suppress regions of the input. This is followed by temporal max pooling, a fully-connected layer and a softmax layer.
    \item \textit{AvastConv}~\cite{DBLP:conf/iclr/KrcalSBJ18}:
this architecture 
comprises of four convolutional layers with kernel sizes equal to 32, 32, 16, and 16, respectively. These convolutional layers are separated by a max pooling layer. After the last convolutional layer, a global average pooling layer is applied. Subsequently, the architecture includes three fully connected layers, culminating in a softmax layer.
\end{itemize}

Using the aforementioned architectures as a basis, 16 malware detectors have been considered:
\begin{itemize}
    \item \textit{NS-MalConv, NS-MalConvGCT, and NS-AvastConv}. These detectors correspond to undefended, non-smoothed, versions of the MalConv, MalConvGCT, and AvastConv  models.
    \item \textit{RS-MalConv, RS-MalConvGCT, and RS-AvastConv}. These detectors implement the randomized ablation smoothing scheme defined by Gibert et al.~\cite{gibert2023_randomizedsmoothing}.
    \item \textit{RsDel-MalConv, RsDel-MalConvGCT, and RsDel-AvastConv}. These detectors implement the randomized deletion smoothing scheme defined by Huang et al.~\cite{huang2023rsdel}.
    
    
    \item \textit{CS-MalConv, CS-MalConvGCT, and CS-AvastConv}. These detectors implement the fixed chunk-based smoothing scheme proposed in Section~\ref{sec:derandomized_smoothing} using MalConv, MalConvGCT, and AvastConv as base detectors.
\end{itemize}

\subsection{Detection Performance on Clean Examples}
\label{sec:clean_performance}

\begin{table*}
    \centering
    \caption{Performance metrics of the non-smoothed and smoothed models on the validation set, the test set and the sub-test of 500 malware examples used for adversarial attack evaluation. With the sub-test set being composed of only malware we simply report the accuracy.}
    \label{tab:nonsmoothed_vs_smoothed_validation_test_performance}
        \begin{tabular}{cccccc}
            \hline
            \multirow{2}{*}{}                             & \multicolumn{2}{c}{Validation set} & \multicolumn{2}{c}{Test set} & Test sub-set\\ \cline{2-6} 
                                                          & Accuracy  & ROC AUC               & Accuracy      & ROC AUC   & Accuracy  \\ \hline
            \multicolumn{1}{l|}{NS-MalConv}     & 96.80  & \multicolumn{1}{c|}{96.74}  & 96.96   & \multicolumn{1}{c|}{96.91}   &  94.20 \\ 
            \multicolumn{1}{l|}{NS-MalConvGCT}  & 95.71  & \multicolumn{1}{c|}{95.65}  & 96.71   & \multicolumn{1}{c|}{96.67}  &  95.00 \\
            \multicolumn{1}{l|}{NS-AvastConv}    & 98.43 & \multicolumn{1}{c|}{98.43}  & 98.69   & \multicolumn{1}{c|}{98.68}   &  97.80 \\ \hline
            \multicolumn{1}{l|}{RS-MalConv}     & 98.77     & \multicolumn{1}{c|}{98.75}  & 98.60   & \multicolumn{1}{c|}{98.57}   &  97.60 \\ 
            \multicolumn{1}{l|}{RS-MalConvGCT}     & 98.22     & \multicolumn{1}{c|}{98.23}  & 99.17  & \multicolumn{1}{c|}{99.17} &  \textbf{99.00} \\ 
            \multicolumn{1}{l|}{RS-AvastConv}     & 98.40     & \multicolumn{1}{c|}{98.39}  & 98.36   & \multicolumn{1}{c|}{98.34}         &  98.40 \\ \hline
            \multicolumn{1}{l|}{RsDel-MalConv}     & 98.17     & \multicolumn{1}{c|}{98.17}  & 98.09   & \multicolumn{1}{c|}{98.07} & 97.40 \\ 
            \multicolumn{1}{l|}{RsDel-MalConvGCT}     & \textbf{99.00}     & \multicolumn{1}{c|}{99.00}  & 99.36   & \multicolumn{1}{c|}{99.36} &  98.80 \\ 
            \multicolumn{1}{l|}{RsDel-AvastConv}     & 98.53    & \multicolumn{1}{c|}{98.51}  & 98.65  & \multicolumn{1}{c|}{98.63}         &  97.40 \\  \hline
        \end{tabular}
    \quad
        \begin{tabular}{cccccc}
            \hline
            \multirow{2}{*}{}                             & \multicolumn{2}{c}{Validation set} & \multicolumn{2}{c}{Test set} & Test sub-set\\ \cline{2-6} 
                                                          & Accuracy  & ROC AUC               & Accuracy      & ROC AUC   & Accuracy  \\ \hline
            \multicolumn{1}{l|}{CS-MalConv (z=512)}     &  \textbf{94.36}  & \multicolumn{1}{c|}{\textbf{94.42}}  & 96.06  & \multicolumn{1}{c|}{96.06}  &  \textbf{96.80}        \\  
            \multicolumn{1}{l|}{CS-MalConv (z=1024)}     &  91.71  & \multicolumn{1}{c|}{91.74}  & 92.90 & \multicolumn{1}{c|}{92.85}  & 91.20 \\ 
            \multicolumn{1}{l|}{CS-MalConv (z=2048)}     &  93.63  & \multicolumn{1}{c|}{93.70}  & 95.64 & \multicolumn{1}{c|}{95.66}  & 96.80 \\  
            \multicolumn{1}{l|}{CS-MalConv (z=4096)}     &  94.36  & \multicolumn{1}{c|}{94.40}  & \textbf{96.23} & \multicolumn{1}{c|}{\textbf{96.22}}  & 96.00 \\  \hline
            \multicolumn{1}{l|}{CS-MalConvGCT (z=512)}  &   94.18     & \multicolumn{1}{c|}{94.24}  & 95.75   & \multicolumn{1}{c|}{95.75}  & 95.80 \\  
            \multicolumn{1}{l|}{CS-MalConvGCT (z=1024)}  &   94.27     & \multicolumn{1}{c|}{94.33}  & \textbf{96.46}  & \multicolumn{1}{c|}{\textbf{96.47}}  & 96.80 \\  
            \multicolumn{1}{l|}{CS-MalConvGCT (z=2048)}  &   \textbf{94.28}     & \multicolumn{1}{c|}{\textbf{94.35}}  & 96.29  & \multicolumn{1}{c|}{96.31} & \textbf{97.00} \\  
            \multicolumn{1}{l|}{CS-MalConvGCT (z=4096)}  &   93.84     & \multicolumn{1}{c|}{93.89}  & 93.26 & \multicolumn{1}{c|}{93.27}  & \textbf{97.00}\\   \hline
            \multicolumn{1}{l|}{CS-AvastConv (z=512)}    &   86.89     & \multicolumn{1}{c|}{86.78}  & \textbf{96.24}  & \multicolumn{1}{c|}{\textbf{96.25}}  & \textbf{95.80} \\  
            \multicolumn{1}{l|}{CS-AvastConv (z=1024)}    &   \textbf{87.08}   & \multicolumn{1}{c|}{\textbf{86.94}}  & 96.05 & \multicolumn{1}{c|}{96.04}  &   94.20 \\  
            \multicolumn{1}{l|}{CS-AvastConv (z=2048)}    &   87.08   & \multicolumn{1}{c|}{86.92}  & 95.85  & \multicolumn{1}{c|}{95.82}  & 93.20 \\  
            \multicolumn{1}{l|}{CS-AvastConv (z=4096)}    &   86.54    & \multicolumn{1}{c|}{86.39}  & 95.10  & \multicolumn{1}{c|}{95.06}  &  92.00    \\  \hline
        \end{tabular}
\end{table*}

Table~\ref{tab:nonsmoothed_vs_smoothed_validation_test_performance} presents the detection accuracy (clean accuracy) and ROC AUC score of the non-smoothed and smoothed-based models on the samples from the validation and test sets. The clean accuracy refers to the ability of the detectors to correctly classify the benign and malicious examples without any adversarial modification or attack employed to manipulate their structure. You can observe that the non-smoothed models (NS-MalConv, NS-MalConvGCT, and NS-AvastConv)
have slightly greater clean accuracy and ROC AUC score ($\approx$2\%) than the (de)randomized smoothing-based models. However, as shown in Section~\ref{sec:evasion_attacks_performance}, the non-smoothed models are vulnerable to evasion attacks that manipulate the structure of the Portable Executable files. In addition, the non-smoothed models do not provide any certification guarantees against the content insertion attacks typically employed by attackers to evade deep learning-based malware detectors.

For the randomized smoothing schemes, we have reduced the number of ablated version from 100 to 20 as it significantly reduces the computational costs without compromising their performance. We refer the readers to Appendix~\ref{appendix:computational_time_mlw_detectors} for a comparison between different model architectures. Furthermore, for the remaining experiments, the probability of ablating a byte and the probability of removing a byte is set to 20\% and to 3\% for the randomized ablation and deletion schemes, respectively, as in their original papers~\cite{gibert2023_randomizedsmoothing,huang2023rsdel}.

\subsection{Certified Accuracy Against Patch and Content Insertion Attacks}
In Section~\ref{sec:derandomized_smoothing}, we proposed a certification procedure to verify and guarantee whether or not the model's predictions will vary in the face of potential adversarial attacks or variations in the input space produced by patch and content insertion attacks. Here, we provide the certified accuracy of the proposed (de)randomized smoothing scheme for the task of malware detection calculated with the aforementioned certification procedure. The certified accuracy is a metric used to quantitatively assess the model's reliability or robustness against a particular type of attack. In our case, we assess the robustness of the malware detectors against patch and content insertion attacks. 

\begin{table*}[ht]
    \caption{Certified accuracy of the smoothing-based detectors against the patch attack. The patch size, $p$, corresponds to a percentage of the original file.}
    \label{tab:certified_accuracy_against_patch_attacks}
    \resizebox{\textwidth}{!}{%
    \begin{tabular}{lcccclllllllll}
    \hline
    \multicolumn{1}{c}{\multirow{2}{*}{}}       & \multicolumn{13}{c}{Certified Accuracy}                                           \\ \cline{2-14} 
    \multicolumn{1}{c}{}                        & $p$=1\% & $p$=2\% & $p$=3\% & $p$=5\% & $p$=10\% & $p$=15\% & $p$=20\% & $p$=25\% & $p$=30\% & $p$=35\% & $p$=40\% & $p$=45\% & $p$=50\% \\ \hline
    \multicolumn{1}{l|}{CS-MalConv (z=512)}   & \textbf{95.36} & \textbf{95.20} & \textbf{95.07} & \textbf{94.77}  & \textbf{93.49}  & \textbf{91.22} & \textbf{88.21}  & \textbf{85.15}  & \textbf{80.46}  & \textbf{74.53}  & \textbf{62.49} & \textbf{40.26}  & 0.0 \\
    \multicolumn{1}{l|}{CS-MalConv (z=1024)}   & 91.55 & 91.33 & 91.18 & 90.91 & 89.64  & 87.74 & 84.71 & 79.25  & 72.68  & 64.76  & 52.86 & 31.73 & 0.0  \\
    \multicolumn{1}{l|}{CS-MalConv (z=2048)}   & 93.88 & 93.42 & 93.37 & 92.89 & 91.63 & 89.71 & 84.98  & 81.59 & 75.88  & 67.63 & 55.93 & 31.73 & 0.0  \\
    \multicolumn{1}{l|}{CS-MalConv (z=4096)}   & 91.96 & 91.90 & 91.68 & 91.51 & 89.77 & 86.15 & 83.52 & 79.48 & 71.59 & 63.13 & 49.06 & 25.34 & 0.0  \\ \hline
    \multicolumn{1}{l|}{CS-MalConvGCT (z=512)}   & 95.18 & 95.08 & 94.81 & 94.27  & 92.86  & 90.78 & 88.63 & 84.27  & 77.83  & 68.84  & 57.58 & 37.69 & 0.0  \\
    \multicolumn{1}{l|}{CS-MalConvGCT (z=1024)}   & \textbf{95.47} & \textbf{95.21} & \textbf{95.08} & \textbf{94.92}  & \textbf{93.78}  & \textbf{92.29} & \textbf{89.22}  & \textbf{85.90}  & \textbf{80.87}  & \textbf{75.73}  & \textbf{65.75} & \textbf{42.37} & 0.0  \\
    \multicolumn{1}{l|}{CS-MalConvGCT (z=2048)}   & 94.25 & 94.03  & 93.93 & 93.67  & 92.64  & 91.14 & 88.22 & 84.06  & 76.62  & 68.05  & 56.87 & 32.44 & 0.0  \\
    \multicolumn{1}{l|}{CS-MalConvGCT (z=4096)}   & 92.83 & 92.54 & 92.48 & 92.18  & 90.66  & 88.27 & 85.16  & 78.26 & 69.62 & 60.16  & 44.73 & 22.55 & 0.0  \\ \hline
    \multicolumn{1}{l|}{CS-AvastConv (z=512)}   & \textbf{95.62} & \textbf{95.45} & \textbf{95.45}  & \textbf{94.93}  & \textbf{94.06}  & \textbf{91.73} & \textbf{88.62}  & \textbf{84.72}  & \textbf{80.64}  & \textbf{73.54} & \textbf{61.55}  & \textbf{39.38}  & 0.0  \\
    \multicolumn{1}{l|}{CS-AvastConv (z=1024)}   & 94.72 & 94.58 & 94.45 & 94.04  & 93.04  & 91.16 & 88.15 & 84.27  & 79.28  & 72.76  & 60.10 & 37.59 & 0.0  \\
    \multicolumn{1}{l|}{CS-AvastConv (z=2048)}   & 94.10 & 93.82 & 93.74 & 93.47  & 92.09  & 89.78 & 87.80  & 84.21  & 80.07  & 71.96  & 59.26 & 31.66- & 0.0  \\
    \multicolumn{1}{l|}{CS-AvastConv (z=4096)}   & 91.60 & 91.42 & 91.21 & 90.08  & 88.73  & 84.33 & 80.88  & 77.00  & 68.86  & 59.54  & 42.08 & 20.99 & 0.0  \\ \hline
    \end{tabular}
    }
\end{table*}

\begin{table*}[ht]
    \caption{Certified accuracy of the smoothed detector against the content insertion attack. The adversarial payload, $p$, corresponds to a percentage of the original file.}
    \label{tab:certified_accuracy_against_content_insertion_attacks}
    \resizebox{\textwidth}{!}{%
    \begin{tabular}{lccccllllllllll}
    \hline
    \multicolumn{1}{c}{\multirow{2}{*}{}}       & \multicolumn{13}{c}{Certified Accuracy}                                           &         \\ \cline{2-15} 
    \multicolumn{1}{c}{}                        & $p$=1\% & $p$=2\% & $p$=3\% & $p$=5\% & $p$=10\% & $p$=20\% & $p$=30\% & $p$=40\% & $p$=50\% & $p$=60\% & $p$=70\% & $p$=80\% & $p$=90\% & $p$=100\% \\ \hline
    \multicolumn{1}{l|}{CS-MalConv (z=512)}   & \textbf{95.83} & \textbf{95.47} & \textbf{95.40} & \textbf{95.27}  & \textbf{94.94}  & \textbf{93.68}  & \textbf{91.57} & \textbf{88.49}  & \textbf{85.57}  & \textbf{81.78}  & \textbf{76.80} & \textbf{65.08}  & \textbf{43.06}  & 0.0   \\
    \multicolumn{1}{l|}{CS-MalConv (z=1024)}   & 92.54 & 92.48 & 92.41 & 92.13  & 91.65  & 90.38  & 88.80  & 85.97  & 81.03  & 75.76  & 67.85 & 56.24  & 34.14  & 0.0    \\
    \multicolumn{1}{l|}{CS-MalConv (z=2048)}   & 95.14 & 95.08 & 94.99 & 94.75  & 94.17  & 93.01  & 91.35  & 88.04  & 84.45  & 79.423  & 73.15 & 64.10  & 37.85  & 0.0    \\
    \multicolumn{1}{l|}{CS-MalConv (z=4096)}   & 94.30 & 94.27 & 94.21 & 94.02  & 93.50  & 92.18  & 89.54  & 87.10  & 83.62  & 79.28  & 72.44 & 58.11  & 32.04 & 0.0     \\ \hline
    \multicolumn{1}{l|}{CS-MalConvGCT (z=512)}   & 95.58 & 95.34 & 95.33 & 95.20  & 94.45  & 93.09  & 91.31  & 89.08  & 84.98  & 80.13 & 71.40 & 59.30  & 39.45  & 0.0    \\
    \multicolumn{1}{l|}{CS-MalConvGCT (z=1024)}   & \textbf{96.27} & \textbf{96.19} & \textbf{96.15} & \textbf{95.89}  & \textbf{95.50}  & \textbf{94.46}  & \textbf{92.86}  & \textbf{89.92}  & \textbf{87.50}  & \textbf{83.37}  & \textbf{78.34} & \textbf{69.36}  & \textbf{48.04}  & 0.0    \\
    \multicolumn{1}{l|}{CS-MalConvGCT (z=2048)}   & 95.77 & 95.58 & 95.47 & 95.18  & 94.85  & 93.91  & 92.43  & 90.01  & 87.20  & 80.95  & 73.91 & 64.25  & 38.54 & 0.0    \\
    \multicolumn{1}{l|}{CS-MalConvGCT (z=4096)}   & 95.10 & 95.07 & 95.07 & 94.93  & 94.40 & 93.24  & 91.29  & 88.63  & 83.90  & 77.27  & 69.82 & 53.55  & 28.89 & 0.0     \\ \hline
    \multicolumn{1}{l|}{CS-AvastConv (z=512)}    & \textbf{96.14} & \textbf{95.86} & \textbf{95.81} & \textbf{95.64}  & \textbf{95.03}  & \textbf{94.23} & 92.13  & 88.97  & 85.49  & 82.41  & 76.24 & 64.38 &  \textbf{41.96} & 0.0 \\
    \multicolumn{1}{l|}{CS-AvastConv (z=1024)}   & 95.59 & 95.57 & 95.51 & 95.37  & 94.75  & 94.07  & \textbf{92.15} & \textbf{89.12}  & 85.80 & 81.76 & 75.99 & 64.50  & 41.21  & 0.0    \\
    \multicolumn{1}{l|}{CS-AvastConv (z=2048)}   & 95.51 & 95.46 & 95.42 & 95.15  & 94.77  & 93.42  & 91.17  & 88.99  & \textbf{86.96}  & \textbf{83.29}  & \textbf{77.54} & \textbf{66.98}  & 40.82  & 0.0    \\
    \multicolumn{1}{l|}{CS-AvastConv (z=4096)}   & 93.98 & 93.94 & 93.90 & 93.52  & 92.57  & 91.33  & 88.35  & 84.66  & 81.40  & 75.76  & 68.88 & 52.35  & 26.94  & 0.0    \\ \hline
    \end{tabular}
    }
\end{table*}

Tables~\ref{tab:certified_accuracy_against_patch_attacks} and~\ref{tab:certified_accuracy_against_content_insertion_attacks} present the certified accuracy of the malware detectors against patches and payloads of different sizes. As Portable Executable files exhibits significant variability in size, we have calculated the certified accuracy of the detectors with respect to adversarial patches and adversarial payloads of different sizes defined as a percentage of the original file's size. 
As you can observe in Tables~\ref{tab:certified_accuracy_against_patch_attacks} and~\ref{tab:certified_accuracy_against_content_insertion_attacks}, the certified accuracy of the detectors diminishes with larger patch and payload sizes until it drops to 0. In the case of patch attacks, the certified accuracy drops to 0 when the adversarial patch matches half the size of the original executable. This decline is attributed to the circumstance where modifying half the size of the executables might lead to having a higher number of chunks mislabelled. In the worst-case scenario, i.e. when the beginning or/and end of the adversarial patch intersects with two chunks instead of one, then half of the chunks plus one will be mislabelled, and thus, the number of chunks that will be correctly labelled will be surpassed, flipping the prediction of the classifier. 
Similarly, in the context of content insertion attacks, the certified accuracy of the detectors reaches 0 when inserting an adversarial payload of the same size as the original executables. In contrast to patch attacks, the content insertion attack does not modify the original content of the executable but adds new content. Consequently, the insertion of a payload matching the original size leads to a situation where the total file size doubles. 

Furthermore, you can observe that the larger the size of the chunks, the smaller the certified accuracy of the detectors. This is due to the fact that the size of the Portable Executable files varies significantly, with files consisting of only a few hundred bytes to 1MB in our experiments. Consequently, the larger the chunk size used to split the files, the fewer the number of chunks. As a result, it is easier to flip the predictions of half of the chunks in a file from the smaller executables. For instance, given a malicious file $x$ of size $L=10000$, if the chunk size is $z=4096$, then the malicious example will be divided into 3 chunks. Considering that all chunks are labelled as malicious, if the attacker modifies an adversarial patch that intersects with two out of the three chunks, and flips their prediction from malicious to benign, then the malware detector will misclassify the malicious example as benign. Notice that in the worst-case scenario, the attacker flips the prediction of a chunk if at least one byte of the adversarial patch intersects with the chunk.

\subsection{Certified Robustness Against State-of-the-Art Evasion Attacks}
\label{sec:evasion_attacks_performance}
\begin{figure}[ht]
    \includegraphics[width=0.5\columnwidth]{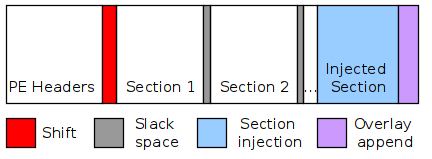}
    \centering
    \caption{Graphical representation of the locations perturbed by different attack strategies.}
    \label{fig:pe_attacks_illustration}
\end{figure}

We now describe the functionality-preserving manipulation attacks, illustrated in Figure~\ref{fig:pe_attacks_illustration}, used to evaluate our defenses. The attacks can be divided into (1) patch attacks and (2) content insertion attacks. Regarding patch attacks, we leverage the Slack~\cite{DBLP:conf/sp/SuciuCJ19} attack that manipulates the already-available space between sections, left by the compiler to maintain the alignment. Regarding content-insertion attacks, we leverage: (i) Padding~\cite{DBLP:conf/eusipco/KolosnjajiDBMGE18} that include new content at the end of the file by just appending bytes as overlay; (ii) Shift~\cite{demetrio2021adversarial} that manipulates the format to inject space before and after each section, later filled with adversarial content; (iii) GAMMA~\cite{demetrio2021functionality} that creates new sections containing chunks extracted from legitimate samples, and (iv) Code-caves~\cite{YUSTE2022102643} that dynamically extend the size of sections in malicious software to inject an adversarial payload. 

The adversarial payloads injected by the aforementioned attacks are optimized with genetic algorithms~\cite{demetrio2021adversarial} to convert the white-box attacks~\cite{YUSTE2022102643,demetrio2021functionality} to black-box attacks, and thus, be able to attack the randomized smoothing-based defenses~\cite{gibert2023_randomizedsmoothing,huang2023rsdel} and our (de)randomized smoothing-based defense. The hyperparameters of the genetic algorithm are the following: (1) the population size is equal to 10, (2) the maximum number of steps is equal to 50, (3) the probability of an individual to be mutation is 30\%, (4) the probability of a byte to be mutated is 30\%, and (5) we use soft label instead of hard label.

As some attacks may require several hours to execute per file, we use a reduced-size test set consisting of 500 malicious examples. By employing a smaller subset, we aim to reduce the computational overhead and accelerate the overall experimentation process. This approach has allowed us to obtain meaningful insights and results within a reasonable timeframe, considering the time-intensive nature of the experiments involved. Nevertheless, we would like to denote that our evaluation set is comparable in size to prior work~\cite{DBLP:conf/eusipco/KolosnjajiDBMGE18,DBLP:conf/sp/SuciuCJ19,demetrio2021adversarial,huang2023rsdel,gibert2023_randomizedsmoothing}.

Following we present
the adversarial accuracy of the baseline detectors and the randomized smoothing-based detectors as well as the certified accuracy (within brackets) of the (de)randomized smoothing-based detectors against the aforementioned SOTA adversarial attacks. 
The adversarial accuracy
refers to the accuracy of the malware detectors against adversarial EXEmples generated with the adversarial attacks. In constrast, the certified accuracy measures the performance of the malware detectors against a worst-case scenario within the specified attacks. In other words, it provides a guarantee of a model's performance against the specified types of manipulations.

\subsubsection{Padding+Slack Space Attack}
\label{sec:padding_attack}
Table~\ref{tab:padding_attack} presents the detection accuracy of the malware detectors against the adversarial EXEmples generated with the padding and slack attack, with adversarial payloads of size 5000 and 10000 bytes. As it can be observed, neither the baseline detectors or the randomized smoothing-based detectors are robust against the padding and slack space attack, even though optimizing the bytes with a genetic algorithm is not as effective as using the Fast Gradient Sign Method (FGSM)~\cite{DBLP:conf/sp/SuciuCJ19}. Nevertheless, you can observe that the attack is able to decrease the accuracy of NS-MalConv, RS-MalConv, and RsDel-MalConv from 94.20, 97.60, and 97.80 to 85.00, 90.00, and 73.80, respectively, by only appending a total of 10000 bytes at the overlay of the adversarial EXEmples. Contrarily, the detection accuracy of the (de)randomized smoothing-based models barely decreases and their certified accuracy is greater than the adversarial accuracy of their corresponding non-smoothed counterparts, highlighting their superior robustness under adversarial conditions. For instance, the certified accuracy of CS-MalConv  is 95.40  while the adversarial accuracy of NS-MalConv is 85.00 when a total payload of 10000 bytes are appended and optimized with a genetic algorithm.

\begin{table}[ht]
    \centering
    \caption{Detection accuracy of the malware detectors on the adversarial EXEmples generated with the padding
+ slack space attack~\cite{DBLP:conf/sp/SuciuCJ19}.}
    \label{tab:padding_attack}
    \begin{tabular}{lccc}
    \hline
    \multirow{2}{*}{Malware detectors} & \multirow{2}{*}{\begin{tabular}[c]{@{}c@{}}Clean \\ examples\end{tabular}} & \multicolumn{2}{c}{Padding+Slack Space Attack }\\ \cline{3-4} 
                                       &                                                                                                  & $p$=5000                       & $p$=10000                      \\ \hline
    NS-MalConv            & 94.20   & 87.40  & 85.00  \\
    RS-MalConv            & 97.60  & 90.40 & 90.00 \\ 
    RsDel-MalConv         & \textbf{97.80} & 82.60 & 73.80 \\ 
    CS-MalConv\,(z=512)    & 96.80 & \textbf{96.20\,(95.80)} & \textbf{95.80\,(95.40)} \\ \hline
    NS-MalConvGCT         & 95.00   & 72.60  & 70.20  \\
    RS-MalConvGCT         & \textbf{99.00}   & \textbf{96.80} & 96.00 \\ 
    RsDel-MalConvGCT      & 98.80 & 81.40 & 68.80 \\ 
    CS-MalConvGCT\,(z=2048)& 97.00  & 96.20\,(94.20) & \textbf{96.20\,(94.40)} \\ \hline
    NS-AvastConv          & 97.80    & \textbf{96.40}  & \textbf{93.60}  \\ 
    RS-AvastConv          & \textbf{98.40}   & 94.60   &  85.60  \\ 
    RsDel-AvastConv       & 97.40   & 93.60   & 89.20   \\ 
    CS-AvastConv\,(z=1024) & 94.20  & 93.60\,(93.20)   & 93.00\,(92.80)   \\ \hline
    \end{tabular}
\end{table}

\subsubsection{Shift Attack}
\label{sec:shift_attack}
Table~\ref{tab:shift_attack} presents the detection accuracy of the malware detectors against the adversarial EXEmples generated by injecting 1024, 2048, and 4096 bytes between the headers and the first section of the adversarial EXEmples. Similar trends are evident in Table~\ref{tab:shift_attack}, mirroring those found in Table~\ref{tab:padding_attack}, i.e. the certified accuracy of the (de)randomized smoothing models is greater than the adversarial accuracy of the baseline and randomized smoothing-based models. For example, the certified accuracy of CS-MalConv, CS-MalConvGCT and CS-AvastConv against the shift attack (extension amount equals to 4096) is 96.60, 96.60 and 93.80, respectively. On the other hand, the adversarial accuracy of the baseline detectors NS-MalConv, NS-MalConvGCT and NS-AvastConv is 35.60, 76.80 and 31.20, respectively.

\begin{table}[ht]
    \centering
    \caption{Detection accuracy of the malware detectors on the adversarial EXEmples generated with the shift attack~\cite{demetrio2021adversarial}. }
    \label{tab:shift_attack}
        \begin{tabular}{lccll}
        \hline
        \multirow{2}{*}{Malware detectors} & \multirow{2}{*}{\begin{tabular}[c]{@{}c@{}}Clean \\ examples\end{tabular}} & \multicolumn{3}{c}{Shift Attack} \\ \cline{3-5} 
                                           &                                                                            & p=1024                       & p=2048                       & p=4096                     \\ \hline
        NS-MalConv      & 94.20     & 42.00    & 35.60     & 35.60\\
        RS-MalConv                & 97.60 & 75.00 & 59.20 & 49.20\\ 
        RsDel-MalConv                & \textbf{97.80} & 86.60 & 80.80 & 77.20\\ 
        \begin{tabular}[c]{@{}c@{}}CS-MalConv \\ (z=512)\end{tabular}      & 96.80     & \textbf{96.80\,(96.80)}    & \textbf{96.80\,(96.60)}      & \textbf{96.60\,(96.60)}      \\ \hline
        NS-MalConvGCT   & 95.00     & 79.40    & 76.00     & 76.80 \\
        RS-MalConvGCT             & \textbf{99.00}  & 95.80  & 94.20 & 81.60 \\ 
        RsDel-MalConvGCT             & 98.80  & 95.40 & 94.20 & 84.80\\ 
        \begin{tabular}[c]{@{}c@{}}CS-MalConvGCT \\ (z=2048)\end{tabular} & 97.00     & \textbf{96.60\,(96.60)}    &  \textbf{96.60\,(96.60)}     &  \textbf{96.60\,(96.60)}      \\ \hline
        NS-AvastConv     & 97.80     & 61.80   & 42.80      & 31.20 \\ 
        RS-AvastConv               & \textbf{98.40} & 64.20  & 63.40 & 52.40 \\ 
        RsDel-AvastConv               & 97.40 & 63.20  & 58.80  & 52.40  \\ 
        \begin{tabular}[c]{@{}c@{}}CS-AvastConv \\ (z=1024)\end{tabular}  & 94.20     & \textbf{94.20\,(94.20)}   & \textbf{94.20\,(94.20)}       &  \textbf{94.20\,(93.80)}      \\ \hline
        \end{tabular}
\end{table}    

\subsubsection{GAMMA Attack}
\label{sec:gamma_attack}
To generate the adversarial EXEmples using GAMMA, we allow the attacker to insert 10 benign sections. In addition, we define the population size of GAMMA equal to 10, the maximum number of steps equal to 50, and we use the soft label. Furthermore, we wanted to limit the maximum amount of injected benign content to  be twice the size of the non-adversarial EXEmple. Hence, we limited the size of each section injected to $\frac{s}{L}$, where $s$ is the total number of sections injected and $L$ is the original size of the EXEmples.

\begin{table}[ht]
    \centering
    \caption{Detection accuracy of the malware detectors on the adversarial EXEmples generated with the GAMMA attack~\cite{demetrio2021functionality}.}
    \label{tab:gamma_attack}
        \begin{tabular}{lcc}
        \hline
        \multirow{2}{*}{Malware detectors} & \multirow{2}{*}{\begin{tabular}[c]{@{}c@{}}Clean \\ examples\end{tabular}} & GAMMA Attack \\ \cline{3-3} 
                                           &                                                                            & p=10   \\ \hline
        NS-MalConv      & 94.20      & 26.20        \\
        RS-MalConv   & 97.60        & 24.20        \\ 
        RsDel-MalConv                & \textbf{97.80} & 91.80 \\ 
        \begin{tabular}[c]{@{}c@{}}CS-MalConv \\ (z=512)\end{tabular}   & 96.80  & \textbf{96.20\,(93.40)}  \\ \hline
        NS-MalConvGCT   & 95.00      & 93.80     \\
        RS-MalConvGCT             & \textbf{99.00}  & \textbf{97.00}  \\ 
        RsDel-MalConvGCT             & 98.80  & 96.80 \\ 
        \begin{tabular}[c]{@{}c@{}}CS-MalConvGCT \\ (z=2048)\end{tabular}  & 97.00 & 93.00\,(84.40)  \\ \hline
        NS-AvastConv     & 97.80      & 58.80     \\ 
        RS-AvastConv               & \textbf{98.40} & 74.20     \\ 
        RsDel-AvastConv               & 97.40 & 74.60 \\ 
        \begin{tabular}[c]{@{}c@{}}CS-AvastConv \\ (z=1024)\end{tabular}    & 94.20 & \textbf{93.60\,(90.60)}  \\ \hline
        \end{tabular}
\end{table}

Table~\ref{tab:gamma_attack} presents the detection accuracy of the malware detectors against the adversarial EXEmples generated with the GAMMA attack. Similar to the results found in Sections~\ref{sec:padding_attack} and~\ref{sec:shift_attack}, the adversarial accuracy, and consequently the certified accuracy, of the chunk-based smoothing classifiers is higher than its non-smoothed counterpart, with the exception of CS-MalConvGCT. For instance, the adversarial accuracy of CS-MalConvGCT is 93.80  while the adversarial accuracy of NS-MalConvGCT is 93.80. As shown in Table~\ref{tab:gamma_attack_CS_models}, this is because of the chunk size of the (de)randomized smoothing-based model, which is equal to 2048 bytes. In general, a smaller chunk size is better against adversarial EXEmples. This is because irrespective of the size of the EXEmples, a smaller chunk size implies that the attacker has to manipulate a greater number of chunks to achieve evasion.

\begin{table}[ht]
\centering
\caption{Comparative analysis of CS-MalConvGCT against adversarial EXEmples generated by the GAMMA attack using a chunk size $z\in\{512, 2048\}$.}
\label{tab:gamma_attack_CS_models}
\begin{tabular}{lcc}
\hline
\multirow{2}{*}{Malware detectors} & \multirow{2}{*}{\begin{tabular}[c]{@{}c@{}}Clean \\ examples\end{tabular}} & GAMMA Attack           \\ \cline{3-3} 
                                   &                                 & $p$=10         \\ \hline
CS-MalConvGCT\,(z=512)               & 95.80                           & \textbf{94.60\,(92.00)}     \\
CS-MalConvGCT\,(z=2048)              & \textbf{97.00}                           & 93.00\,(84.40) \\ \hline
\end{tabular}%
\end{table}

\subsubsection{Code Caves Optimization Attack}
\label{sec:code_caves_attack}
Table~\ref{tab:code_caves_optimization_attack} presents the detection accuracy of the malware detectors against the adversarial EXEmples generated by the code caves optimization attack. Similarly to Sections~\ref{sec:padding_attack},~\ref{sec:shift_attack} and~\ref{sec:gamma_attack}, the adversarial and certified accuracy of the (de)randomized smoothing models is greater than that of the baseline and randomized smoothing-based models, demonstrating the superior robustness of our defense against the code caves attack. For instance, the certified accuracy of CS-MalConv is 90.20 while the adversarial accuracy of NS-MalConv, RS-MalConv and RsDel-MalConv is 22.20, 25.20, and 68.60, respectively. 

\begin{table}[ht]
    \centering
    \caption{Detection accuracy of the malware detectors on the adversarial EXEmples generated with the code caves optimization attack~\cite{YUSTE2022102643}.}
    \label{tab:code_caves_optimization_attack}
    \begin{tabular}{lcc}
        \hline
        Malware detectors      & Clean examples & Code Caves Attack \\ \hline
        NS-MalConv             & 94.20 & 22.20\\
        RS-MalConv             & 97.60 & 25.20\\ 
        RsDel-MalConv          & \textbf{97.80} & 68.60\\ 
        CS-MalConv\,(z=512)    & 96.80             & \textbf{96.00\,(90.40)}                \\ \hline
        NS-MalConvGCT          & 95.00 & 62.60\\
        RS-MalConvGCT          & \textbf{99.00} & 82.40\\ 
        RsDel-MalConvGCT          & 98.80 & 58.00\\ 
        CS-MalConvGCT\,(z=2048)  & 97.00              & \textbf{95.40\,(84.00)}                \\ \hline
        NS-AvastConv            & 97.80 & 20.60\\ 
        RS-AvastConv            & \textbf{98.40} & 40.00\\ 
        RsDel-AvastConv            & 97.40 & 36.80\\
        CS-AvastConv\,(z=1024)    & 94.20              & \textbf{93.00\,(86.20)}                 \\ \hline        
    \end{tabular}
\end{table}

\subsubsection{Combining All Attacks Into One}
In this section we attack the ML-based malware detectors with an attack that combines all the aforementioned manipulations into a single attack. The attack consists of five main steps: (1) shift content between the header and the sections by 4096 bytes, (2) extend code caves between sections, (3) create five new sections filled with benign content, (4) append 10000 bytes in the overlay, and (5) optimize the content with a genetic algorithm. As observed in Table~\ref{tab:combination_attack}, the adversarial accuracy and certified accuracy of the (de)randomized smoothing models is greater than that of their non-smoothing and randomized smoothing counterparts, showing greater resilience against complex evasion attacks. 

\begin{table}[ht]
    \centering
    \caption{Detection accuracy of the malware detectors on the adversarial EXEmples generated with the combination attack.}
    \label{tab:combination_attack}
    \begin{tabular}{lcc}
        \hline
        Malware detectors      & Clean examples & Adversarial examples \\ \hline
        NS-MalConv             & 94.20 & 33.00\\
        RS-MalConv             & 97.60 & 25.00\\ 
        RsDel-MalConv             & \textbf{97.80} & 66.00\\ 
        CS-MalConv\,(z=512)     & 96.80             & \textbf{85.20\,(81.60)}               \\ \hline
        NS-MalConvGCT          & 95.00 & 81.00\\
        RS-MalConvGCT          & \textbf{99.00} & 75.20\\ 
        RsDel-MalConvGCT          & 98.80 & 66.80\\ 
        CS-MalConvGCT\,(z=2048)  & 97.00              & \textbf{83.00\,(69.60)}               \\ \hline
        NS-AvastConv            & 97.80 & 31.20\\ 
        RS-AvastConv            & \textbf{98.40} & 33.40\\ 
        RsDel-AvastConv            & 97.40 & 29.20\\ 
        CS-AvastConv\,(z=1024)    & 94.20              & \textbf{79.20\,(73.00)}                \\ \hline
    \end{tabular}
\end{table}

\subsection{Visualization Tool}
\begin{figure}[ht]
\centering
\begin{subfigure}{.5\textwidth}
  \centering
  \includegraphics[width=.9\linewidth]{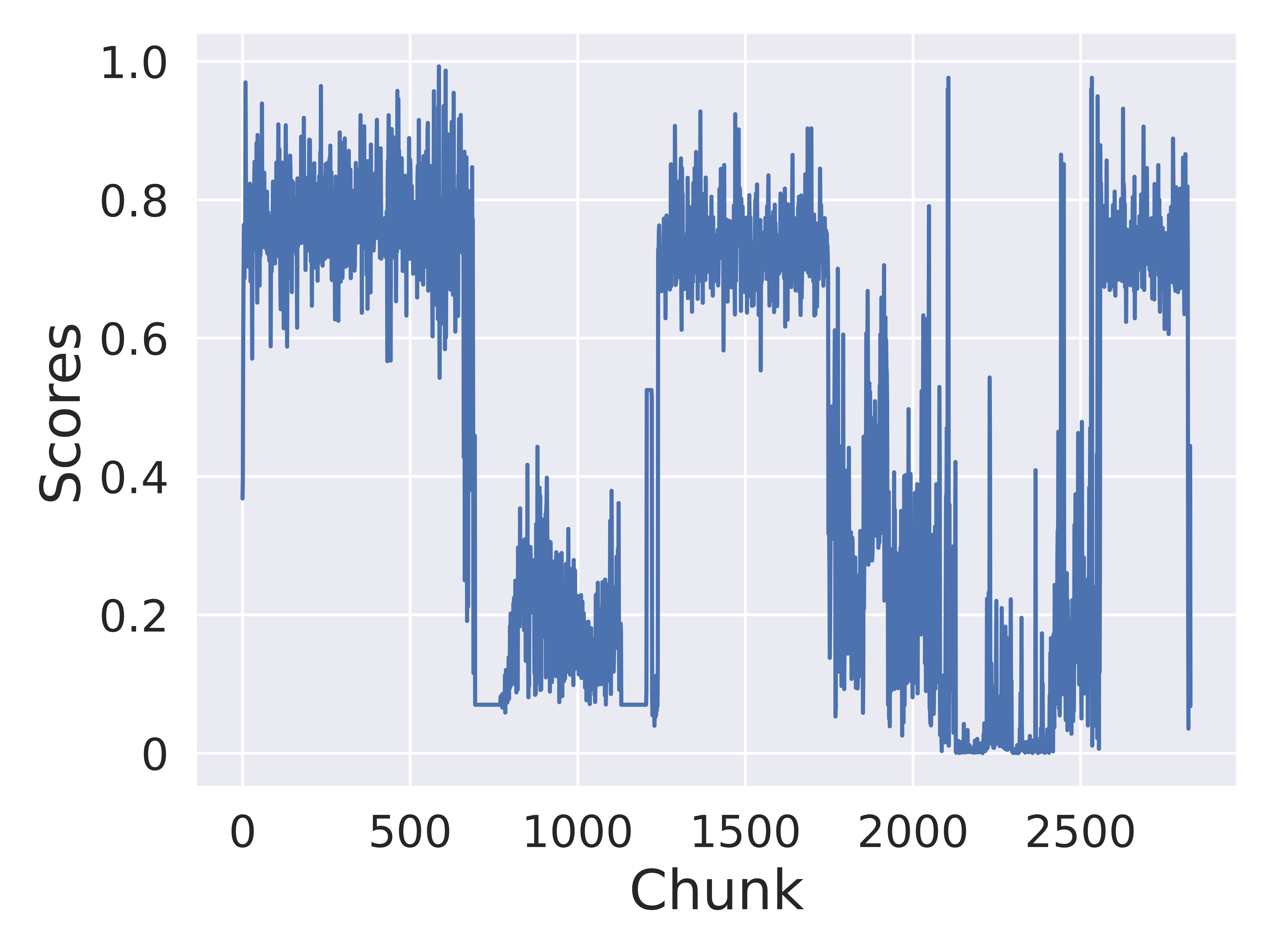}
\end{subfigure}%
\begin{subfigure}{.5\textwidth}
  \centering
  \includegraphics[width=.9\linewidth]{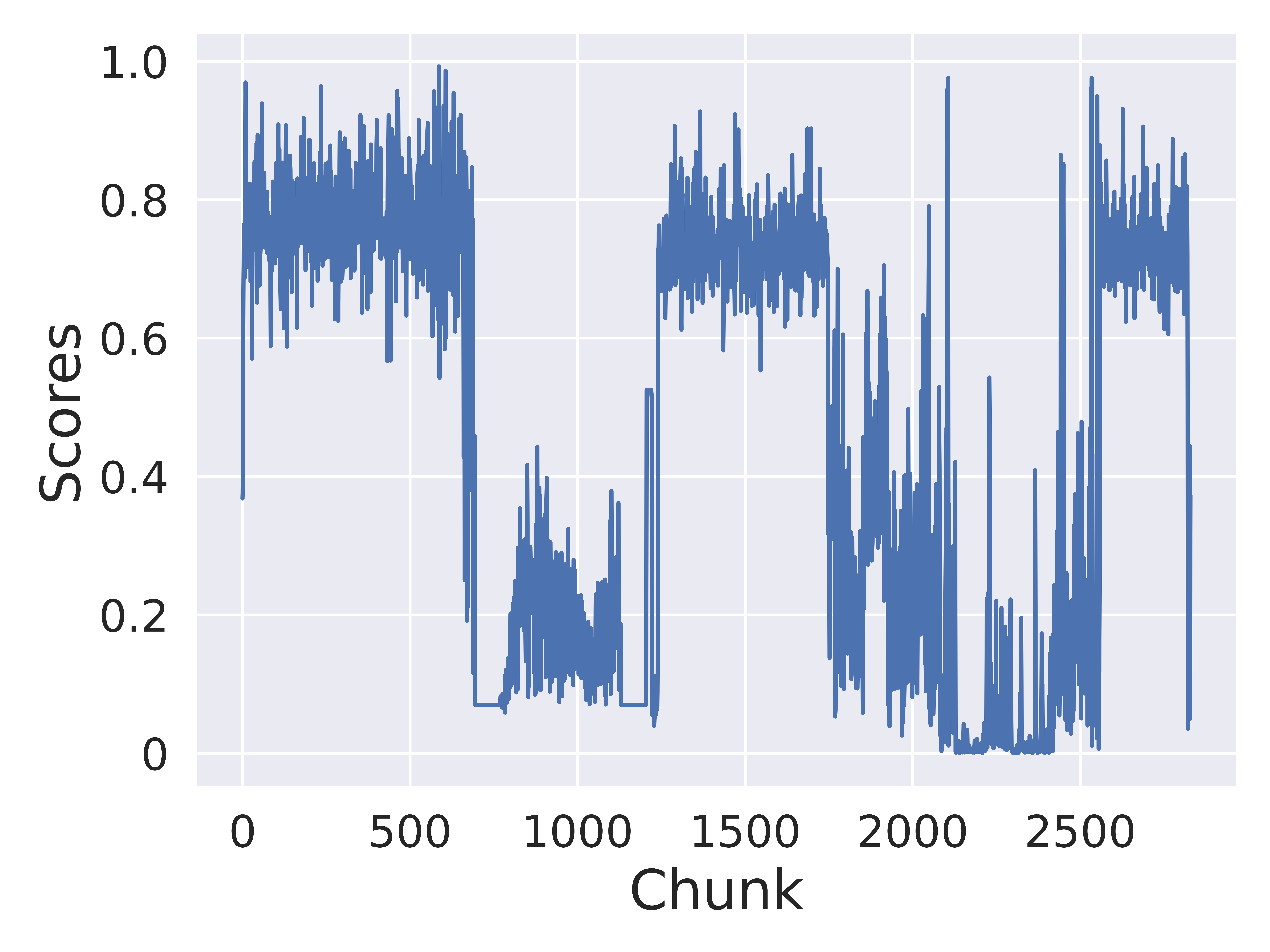}
\end{subfigure}
\caption{Graphical visualization of the scores of EXEmples belonging to the Autoinject malware family.}
    \label{fig:scores_autoinject_family}
\end{figure}

\begin{figure}[ht]
\centering
\begin{subfigure}{.5\textwidth}
  \centering
  \includegraphics[width=.9\linewidth]{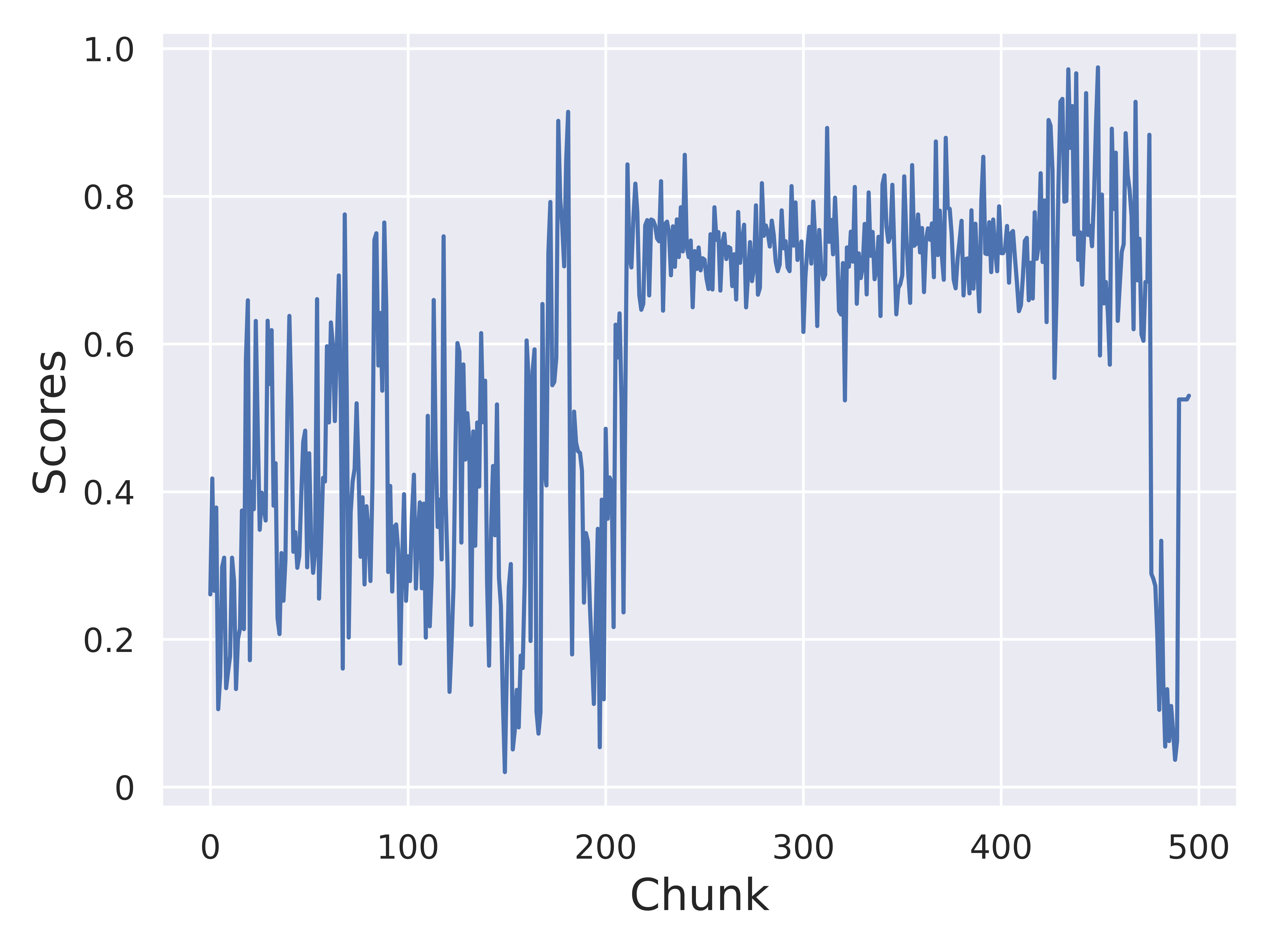}
\end{subfigure}%
\begin{subfigure}{.5\textwidth}
  \centering
  \includegraphics[width=.9\linewidth]{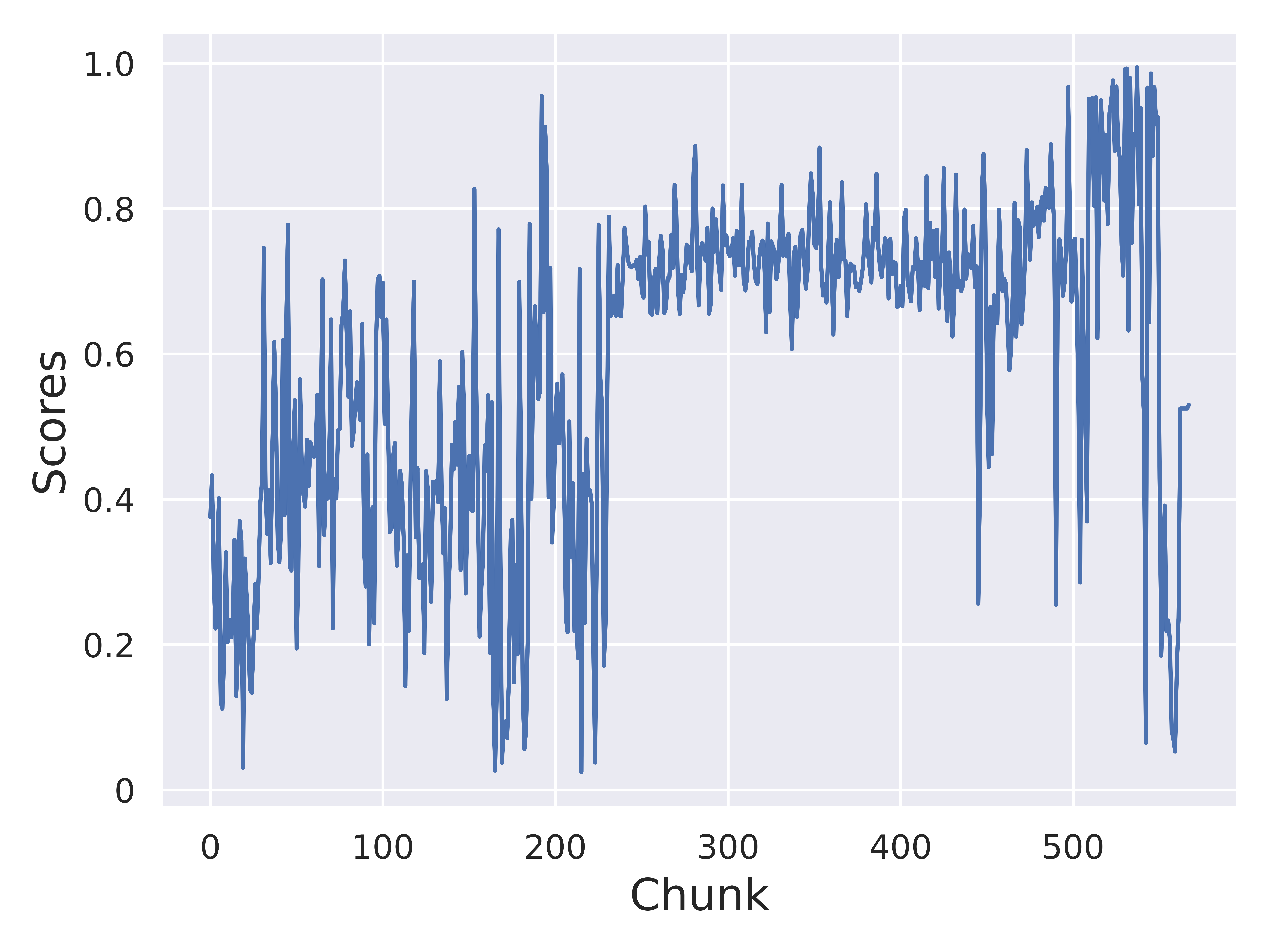}
\end{subfigure}
\caption{Graphical visualization of the scores of EXEmples belonging to the GandCrab malware family.}
    \label{fig:scores_autoit_family}
\end{figure}

The proposed chunk-based smoothing classification system is interpretable by design. This is because the proposed chunk-based smoothing classification system assesses the maliciousness of each chunk independently, facilitating a finer-grained analysis of the file. This allows us to identify which specific chunks within a file exhibit malicious or benign traits. As a result, our approach could be used to help cybersecurity analysts to identify which parts of the suspicious file contains the malicious code. Furthermore, the predictions for each chunk can serve as the basis for generating a visual depiction or an overview of the file, emphasizing chunks with the higher and lower maliciousness scores. 
For instance, Figures~\ref{fig:scores_autoinject_family}, and~\ref{fig:scores_autoit_family} provide a graphic depiction of the scores of examples beloning to the AutoInject and GandCrab families of malware. You can observe that the visualization of EXEmples belonging to the same family are similar between them while distinct from the EXEmples belonging to different families.

\section{Discussion}
\label{sec:discussion}
In this section we discuss the strengths and weaknesses of the proposed (de)randomized smoothing approach with respect to the baseline classifiers (NS) and the randomized smoothing classifiers (RS, RsDel). 

\begin{figure}[ht]
    \centering
    \includegraphics[width=0.60\columnwidth]{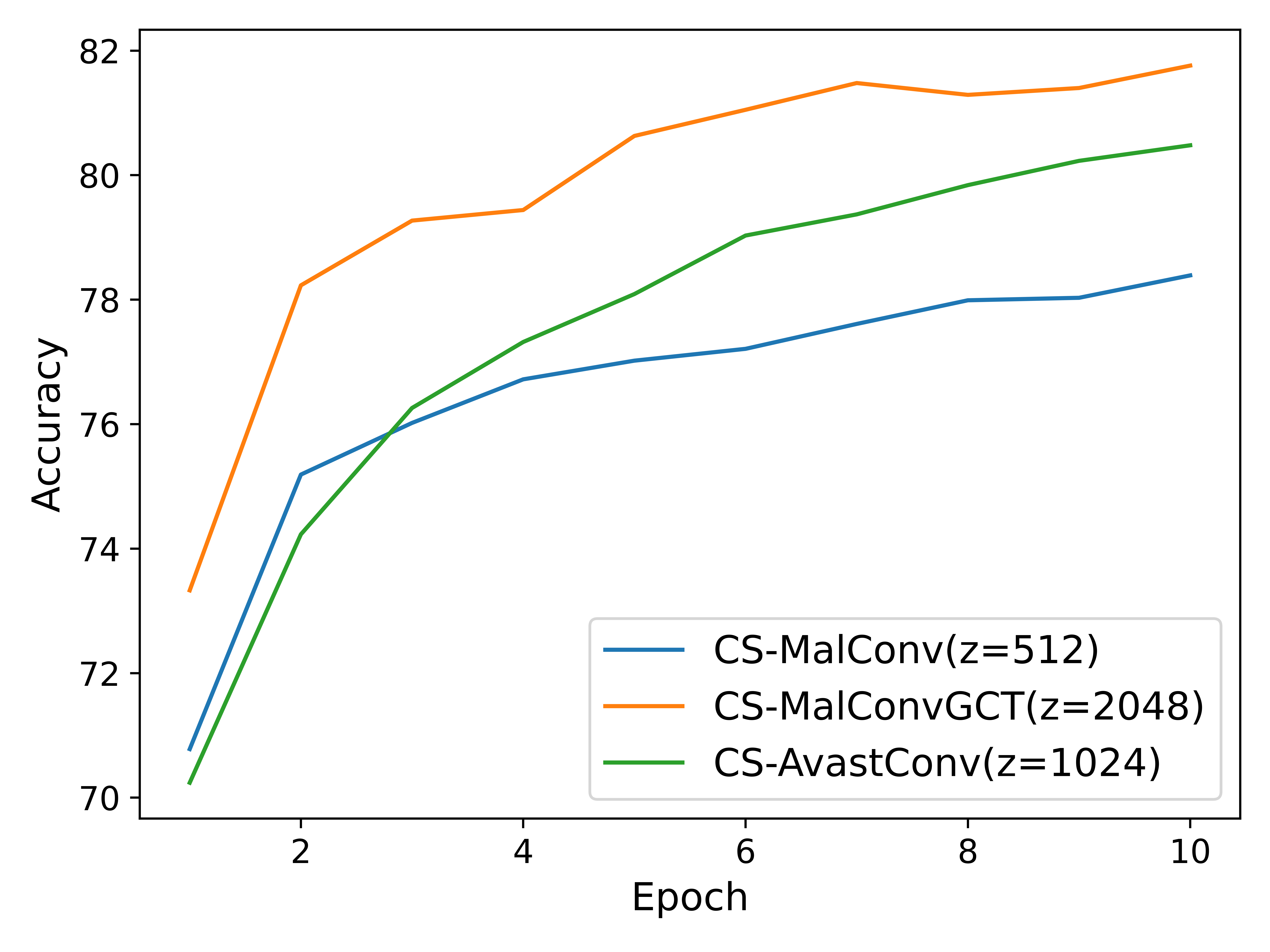}
    \caption{Training accuracy of the (de)randomized smoothing classifiers. The training accuracy is
calculated based on a single chunk per file.}
    \label{fig:train_acc_lineplot}
\end{figure}

By training an end-to-end malware detector using as input the whole bytes of an executable (NS, RS, RsDel), the classifier has access to all the available information in an executable, and thus, can learn to recognize patterns related to the overall structure of the binary file, including the headers, the import address and export address tables, the section tables, or specific data blocks. However, as it has been shown in the literature~\cite{DBLP:conf/itasec/DemetrioBLRA19}, the baseline classifiers~\cite{DBLP:conf/aaai/RaffBSBCN18} tend to learn features based on characteristics of the PE headers that are not necessarily benign or malicious, i.e. artifacts found in the training data, instead of meaningful characteristics from the content of the text and data sections. The randomized smoothing approaches, on the other hand, introduce a randomization step during training, which involve removing~\cite{huang2023rsdel} or ablating~\cite{gibert2023_randomizedsmoothing} certain bytes. This randomization step, as shown in Section~\ref{sec:evasion_attacks_performance}, makes the randomized smoothing classifiers slightly more robust against adversarial attacks. Nevertheless, training an end-to-end malware classifier using the whole byte sequences as input is computationally demanding, as the input of the model might be in the order of millions of bytes. For
instance, the original MalConv model~\cite{DBLP:conf/aaai/RaffBSBCN18}, constrained to binary files up to 2Mb, was trained using data parellelism accross the 8 GPUs of a DGX-1 server, whose price is beyond the means of most research groups’ budgets. Fortunately for researchers, there has been a decrease in computing costs over the last years, allowing research groups with budget constraints to train their own shallow end-to-end models~\cite{GIBERT2021102159}. Regarding the (de)randomized smoothing classifiers, the computational cost of training a chunk-based classifier is minimal as it only requires to process a small portion of the input file, i.e. 512, 1024, 2048 bytes. In addition, training on a subset of bytes inherently limits the amount of information the model can use for classification, preventing the model from learning high-level features from PE files, i.e. features related to the overall PE structure. However, while beneficial, training using a single subset of bytes results in a loss of information with the subsequent reduction in the accuracy of the chunk-based classifier as shown in Figure~\ref{fig:train_acc_lineplot}. Nonetheless, the reduced accuracy on individual chunks is greatly mitigated by the majority voting strategy used by the (de)randomized smoothing approaches as shown in Section~\ref{sec:clean_performance}.

\begin{figure}[ht]
    \centering
    \includegraphics[width=0.60\columnwidth]{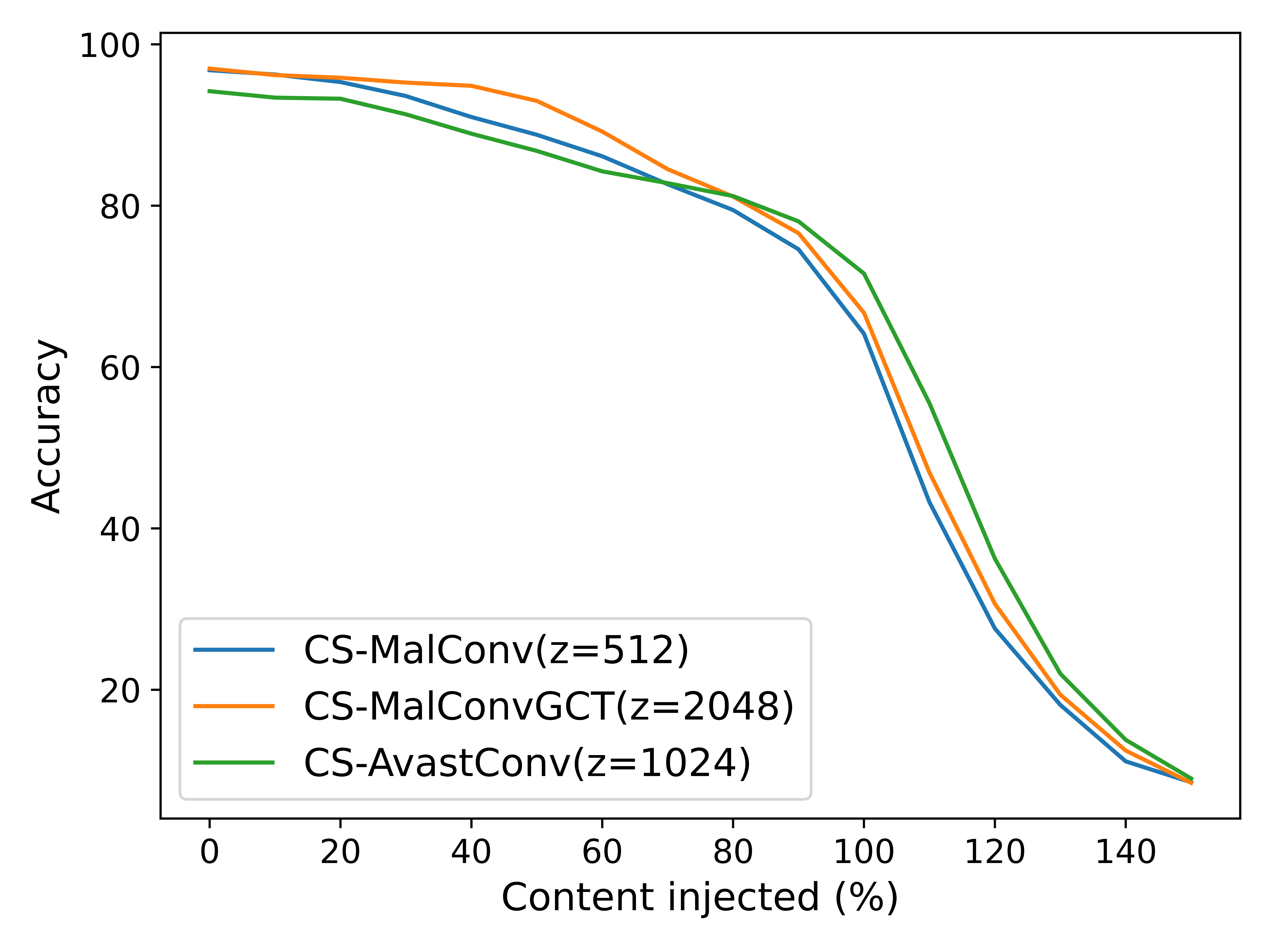}
    \caption{Detection accuracy of the (de)randomized smoothing classifiers against the benign content injection attack with payloads of various sizes.}
    \label{fig:benign_injection_lineplot}
\end{figure}

At inference time, the baseline classifiers are more efficient as no preprocessing step is applied and they only require assessing a given input sample once in comparison to the randomized smoothing classifiers (RS and RsDel), which generate and classify $N$ noisy versions of a given input sample. For a comparative analysis of the inference time of the different types of classifiers we refer the reader to Appendix~\ref{appendix:computational_time_mlw_detectors}. Although faster, taking as input the whole binary file makes the baseline classifiers vulnerable to adversarial attacks~\cite{DBLP:conf/sp/SuciuCJ19,demetrio2021adversarial,demetrio2021functionality,YUSTE2022102643}. On the other hand, randomized smoothing classifiers defend slightly better against adversarial EXEmples although their increased inference time renders those classifiers unfeasible for real-time detection. In contrast, the (de)randomized smoothing classifiers (CS) have exhibited unmatched robustness against SOTA adversarial attacks as the adversarial manipulations won't affect all the chunks but only a subset of them, surpassing randomized smoothing-based classifiers. Unfortunately, a knowledgeable attacker may be able to achieve evasion by exploiting the nature of the (de)randomized smoothing process. Specifically, this process involves splitting an executable into non-overlapping chunks, independently classifying each chunk, and determining the final classification output through a majority voting mechanism. As a result, an attacker could craft an evasive adversarial EXEmple by injecting a huge adversarial payload of size comparable to the original malicious example. By doing so, the adversarial payload has the potential to manipulate the majority voting mechanism, leading to a classification outcome where more than half of the chunks of the adversarial EXEmple are classified as benign, and thus, the adversarial EXEmple will end up being erroneously classified as benign. To assess the performance of the (de)randomized smoothing approach against huge adversarial payloads, Figure~\ref{fig:benign_injection_lineplot} illustrates the detection accuracy of the (de)randomized smoothing classifiers against the benign code injection attack with payloads of various sizes. As it can be observed, if an attacker is able to inject payloads equal or greater to 100\% of the original file's size, then the detection accuracy of the classifiers significantly diminish, highlighting a key vulnerability of our (de)randomized smoothing scheme. 

Nevertheless, our (de)randomized smoothing scheme forces the attackers to inject substantial adversarial payloads to successfully evade detection. This stands in stark contrast to the baseline and the randomized smoothing-based classifiers, which can be easily circumvented with small payloads consisting of just a few hundreds of bytes\,\cite{DBLP:conf/sp/SuciuCJ19,demetrio2021adversarial}.

\section{Conclusions}
\label{sec:conclusions}
In this paper we present the first model-agnostic certified defense against functionality-preserving adversarial attacks on end-to-end malware detectors. Building upon recent research on (de)randomized smoothing, we propose a chunk-based smoothing scheme that provides certified robustness guarantees against patch and content insertion attacks. We rigorously evaluate our certified defense using three neural network architectures for malware detection, (1) MalConv, (2) MalConvGCT, and (3) AvastConv, demonstrating superior robustness against a plethora of adversarial attacks. The novel application of our defense mechanism establishes a new benchmark in terms of detection accuracy against adversarial EXEmples and achieves a significant advancement in the field. 
Our findings suggest a number of directions for future research. The most apparent direction identified involves adapting and integrating recent defenses from the Computer Vision domain for the task of malware detection. Additionally, a line of research could be the exploration of techniques aimed at identifying and removing the adversarial content from the adversarial EXEmples.

\section*{Acknowledgements}
This project has received funding from the European Union’s Horizon 2020 Research and Innovation programme under the Marie Skodowska-Curie Grant Agreement”(as per Article 29.4 of the grant agreement). 
We would like to thank Cormac Doherty and UCD's Centre for Cybersecurity and Cybercrime Investigation for their support.
This project has been also partially support by Fondazione di Sardegna under the project ``TrustML: Towards Machine Learning that Humans Can Trust’’, CUP: F73C22001320007; and European Union's Horizon Europe research and innovation program under the project ELSA, grant agreement No 101070617; and European Union’s Horizon 2020 research and innovation programme under project TESTABLE, grant agreement No 101019206; and SERICS (PE00000014) under the MUR National Recovery and Resilience Plan funded by the European Union – NextGenerationEU, and by MCIN/AEI/10.13039/501100011033/FEDER,~UE under the project PID2022-139835NB-C22.

\bibliographystyle{unsrt}  
\bibliography{references}  

\appendix

\section*{Proofs}
\label{sec:proofs}
We define the Proof using a rationale akin to that presented by Levine et al.~\cite{DBLP:conf/nips/0001F20a}. 
Recall the definitions and statement of Theorem 1. In
particular, recall the base classification counts $n_c(x)$:
\begin{equation}
    \forall_c \in C, n_c(x) \coloneqq \sum_{b=0}^{N} f_c (\tilde{x_b}) \tag{1}
\end{equation} 
And recall the definition of the smoothed classifier:
\begin{equation}
    g(x) := \underset{c}{\arg\max}\: n_c(x) \tag{2}
\end{equation}
where in the case of ties, we choose the larger-indexed class as the argmax solution, i.e., malicious class.

\begin{mythm}{1}
\label{theorem:certification_patch_attack}
    For any input example $x$, base classifier $f$, smoothing classifier $g$, smoothing chunk size $z$ and patch size $p$, such as $\Delta = \ceil*{\frac{p}{z}}+1$, if:
    \begin{equation}
       n_{c^{'}}(x) \ge  \underset{c^{''} \neq c^{'}}{\max}  \left[n_{c^{''}}(x) +1_{c^{'}<c^{''}}\right]+ 2\Delta \tag{4}
    \end{equation}
    where $n_{c^{'}}(x)$ and $n_{c^{''}}(x)$ denote the most commonly and second most commonly predicted classes for an input example $x$, then for any adversarial EXEmple $x^{'}$ which differs from $x$ only in a patch of size $p$, $g(x^{'})=c^{'}$.
\end{mythm}

\begin{proof}
Recall that an input example $x$ of size $L$ is split into contiguous, non-overlapping chunks of size $z$, $x = \{\tilde{x_0}, \tilde{x_1}, \tilde{x_2}, \ldots, \tilde{x_N} \}$, where $N = \ceil*{\frac{L}{z}}$ and $\tilde{x_b}$ represent a chunk of bytes from $x$ starting and ending at positions $b\times z$ and $(b+1)\times z$, respectively.

Let $k$ represent the starting byte or position within $x^{'}$ where the patch is inserted. Then, the end byte will be $k+p$, where $p$ is the size of the patch. Note that, for all $c\in C=\{c_1, c_2\}$, the output $f_c(\tilde{x_b})$ will be equal to $f_c(\tilde{x}^{'}_b)$, unless the chunk $\tilde{x}_b$ intersects with the adversarial patch of size $p$ starting at position $k$ and ending at position $k+p$. This condition occurs only when $\tilde{x}_b$ is in the range $k$ and $k+p$. Notice that there will be at most $\Delta = \ceil*{\frac{p}{z}}+1$ chunks that will meet this condition. Therefore, $f_c(\tilde{x_b}) = f_c(\tilde{x}^{'}_b)$ in all but $\Delta$ cases and thus, because $f_c(\cdot )\in \{0, 1\}$,

\begin{equation}
    \label{eq:proof}
    \forall n_c, \left|n_c(x) - n_c(x^{'})\right| \le \Delta.
\end{equation}

Now consider any $c^{''}\neq c^{'}$ such that $n_{c^{'}}(x) \ge  \left[n_{c^{''}}(x) +1_{c^{'}<c^{''}}\right]+ 2\Delta$. There are two cases:

\begin{itemize}
    \item $c^{'} < c^{''}$: In this case, in the event that $n_{c^{'}}(x^{'})= n_{c^{''}}(x^{'})$, we have that $g(x^{'}) = c^{''}$. Therefore, a sufficient condition for $g(x^{'}) \neq c^{''}$ is that $n_{c^{'}}(x^{'})> n_{c^{''}}(x^{'})$. By Equation~\ref{eq:proof} and triangle inequality, this must be true if  $n_{c^{'}}(x) > \left[ n_{c^{''}}(x) \right] +2\Delta$, or equivalently, if $n_{c^{'}}(x) \ge \left[ n_{c^{''}}(x) + 1_{c^{'}<c^{''}}\right] +2\Delta$.
    \item $c^{''} < c^{'}$: In this case, in the event that $n_{c^{'}}(x^{'})= n_{c^{''}}(x^{'})$, we have that $g(x^{'}) = c^{'}$. Therefore, a sufficient condition for $g(x^{'}) \neq c^{''}$ is that $n_{c^{'}}(x^{'})\ge n_{c^{''}}(x^{'})$. By Equation~\ref{eq:proof} and triangle inequality, this must be true if $n_{c^{'}}(x) \ge  \left[ n_{c^{''}}(x) + 1_{c^{'}<c^{''}}\right] +2\Delta$.
\end{itemize}
Therefore, if $n_{c^{'}}(x) \ge \underset{c^{''} \neq c^{'}}{\max} \left[n_{c^{''}}(x) +1_{c^{'}<c^{''}}\right]+ 2\Delta$, then no class other than $c$ can be output by $g(x^{'})$. 
\end{proof}

The proof of Theorem~\ref{theorem:certification_append_attack}  can be conducted in a similar manner by replacing $2\Delta$ with $\Delta$ as follows:

\begin{mythm}{2}
    \label{theorem:certification_append_attack}
    For any input example $x$, base classifier $f$, smoothing classifier $g$, smoothing chunk size $z$ and adversarial payload of size $p$, such as $\Delta = \ceil*{\frac{p}{z}}+1$, if:
    \begin{equation}
       n_{c^{'}}(x) \ge \underset{c^{''} \neq c^{'}}{max} \left[n_{c^{''}}(x) +1_{c^{'}<c^{''}}\right]+ \Delta
    \end{equation}
    where $n_{c^{'}}(x)$ and $n_{c^{''}}(x)$ denote the most commonly and second most commonly predicted classes for an input example $x$, then for any adversarial EXEmple $x^{'}$ which differs from $x$ only in an adversarial payload of size $p$, $g(x^{'})=c^{'}$.
\end{mythm}

\begin{proof}
Let $k$ represent the starting byte or position within $x^{'}$ where the adversarial payload is inserted. Then, the end byte will be $k+p$, where $p$ is the size of the adversarial payload. Note that, for all $c\in C=\{c_1, c_2\}$, the output $f_c(\tilde{x_b})$ will be equal to $f_c(\tilde{x}^{'}_b)$, unless the chunk $\tilde{x}_b$ intersects with the adversarial patch of size $p$ starting at position $k$ and ending at position $k+p$. Due to the preprocessing step introduced in Section~\ref{sec:insertion_attacks}, this holds true as long as the attacker inserts the adversarial payload at the end of existing sections, between sections, in newly-created sections or at the overlay. As a result, there will be at most one chunk from $x^{'}$ that will intersect with the adversarial payload and $\ceil*{\frac{p}{z}}$ chunks containing the newly-inserted adversarial payload. Therefore, $f_c(\tilde{x_b}) = f_c(\tilde{x}^{'}_b)$ in all but $\Delta =\ceil*{\frac{p}{z}}+1$ cases and thus, because $f_c(\cdot )\in \{0, 1\}$,

\begin{equation}
    \label{eq:proof}
    \forall n_c, \left|n_c(x) - n_c(x^{'})\right| \le \Delta.
\end{equation}

Now consider any $c^{''}\neq c^{'}$ such that $n_{c^{'}}(x) \ge  \left[n_{c^{''}}(x) +1_{c^{'}<c^{''}}\right]+ \Delta$. There are two cases:

\begin{itemize}
    \item $c^{'} < c^{''}$: In this case, in the event that $n_{c^{'}}(x^{'})= n_{c^{''}}(x^{'})$, we have that $g(x^{'}) = c^{''}$. Therefore, a sufficient condition for $g(x^{'}) \neq c^{''}$ is that $n_{c^{'}}(x^{'})> n_{c^{''}}(x^{'})$. By Equation~\ref{eq:proof} and triangle inequality, this must be true if  $n_{c^{'}}(x) > \left[ n_{c^{''}}(x) \right] +\Delta$, or equivalently, if $n_{c^{'}}(x) \ge \left[ n_{c^{''}}(x) + 1_{c^{'}<c^{''}}\right] +\Delta$.
    \item $c^{''} < c^{'}$: In this case, in the event that $n_{c^{'}}(x^{'})= n_{c^{''}}(x^{'})$, we have that $g(x^{'}) = c^{'}$. Therefore, a sufficient condition for $g(x^{'}) \neq c^{''}$ is that $n_{c^{'}}(x^{'})\ge n_{c^{''}}(x^{'})$. By Equation~\ref{eq:proof} and triangle inequality, this must be true if $n_{c^{'}}(x) \ge  \left[ n_{c^{''}}(x) + 1_{c^{'}<c^{''}}\right] +\Delta$.
\end{itemize}
Therefore, if $n_{c^{'}}(x) \ge \underset{c^{''} \neq c^{'}}{\max} \left[n_{c^{''}}(x) +1_{c^{'}<c^{''}}\right]+ \Delta$, then no class other than $c$ can be output by $g(x^{'})$. 
\end{proof}

\section*{Inference Time of the End-to-End Malware Detectors}
\label{appendix:computational_time_mlw_detectors}
We analyse the time required by all the models we have trained to quantify the toll imposed by our certification approach.
As shown in Table~\ref{tab:inference_time}, the baseline architecture that is more efficient, independently of their defense scheme, is MalConv. This is mainly because of two reasons: (1) MalConv uses very large strides compared to AvastConv; (2) MalConv makes use of the Temporal Max-Pooling strategy presented by Raff et al.~\cite{DBLP:conf/aaai/RaffFZAFM21}, which leverage sparse gradients to significantly reduce the runtime and memory constraints; (3) MalConv employs feed-forward layers instead of an Attention layer as in MalConvGCT.
In contrast, the computational time of the randomized smoothing-based models significantly surpasses that of the baseline and (de)randomized smoothing models.

\begin{table}[ht]
    \centering
    \caption{Inference time of the malware detectors (seconds/example).} 
    \label{tab:inference_time}
    \begin{tabular}{lcc}
        \hline
        Malware Detector      & CPU Inference Time & GPU Inference Time\\ \hline
        NS-MalConv               & \textbf{0.0132} & \textbf{0.0213}\\
        NS-MalConvGCT            & 0.0579          & 0.0093 \\
        NS-AvastConv             & 0.0542          & 0.0122 \\ \hline
        RS-MalConv (N=20)        & \textbf{0.2882} & \textbf{0.0287}\\ 
        RS-MalConvGCT (N=20)     & 1.4038          & 0.0404 \\ 
        RS-AvastConv (N=20)      & 1.4075          & 0.0309 \\  \hline
        RsDel-MalConv (N=20)     & \textbf{0.2736} & \textbf{0.0275}\\ 
        RsDel-MalConvGCT (N=20)  & 1.3270          & 0.0408 \\ 
        RsDel-AvastConv (N=20)   & 1.2751          & 0.0299 \\ \hline
        CS-MalConv (z=512)       & \textbf{0.0543} & 0.0241  \\
        CS-MalConvGCT (z=2048)   & 0.5537          & 0.0414 \\
        CS-AvastConv (z=1024)    & 0.6734          & \textbf{0.0220} \\ \hline

    \end{tabular}
\end{table}

Furthermore, Table~\ref{tab:comparative_analysis_N_ablated_versions} provides a comparative analysis between the randomized smoothing-based models employing 20 and 100 ablated versions. 
Results suggests that using 100 ablated versions demonstrates marginal differences in performance, greatly increasing the inference time.
For this reason, we used 20 ablated versions in our experiments.

\begin{table}[ht]
\centering
\caption{Comparative analysis of randomized smoothing-based approaches with different number of ablated versions $N\in\{20, 100\}$.}
    \label{tab:comparative_analysis_N_ablated_versions}
\resizebox{\textwidth}{!}{%
\begin{tabular}{lcccccc}
\hline
\multirow{3}{*}{Malware detectors} & \multicolumn{6}{c}{Randomized Smoothing}                                                                                                                                                                                                                                                                                                                                                                \\ \cline{2-7} 
                                   & \multicolumn{3}{c}{N=20}                                                                                                                                                                           & \multicolumn{3}{c}{N=100}                                                                                                                                                                          \\ \cline{2-7} 
                                   & ROC AUC                        & \begin{tabular}[c]{@{}c@{}}CPU inference time \\ (seconds/example)\end{tabular} & \begin{tabular}[c]{@{}l@{}}GPU inference time\\ (seconds/example)\end{tabular} & ROC AUC                        & \begin{tabular}[c]{@{}c@{}}CPU inference time \\ (seconds/example)\end{tabular} & \begin{tabular}[c]{@{}l@{}}GPU inference time\\ (seconds/example)\end{tabular} \\ \hline
RS-MalConv                         & \textbf{98.75} & \textbf{0.2882}         & \textbf{0.0287}                                                                         & \textbf{98.70} & \textbf{1.5536}                                                & \textbf{0.1564}                                                                         \\
RS-MalConvGCT                      & 98.23                           & 1.4038                                                                          & 0.0404                                                                         & 98.32                           & 7.9180                                                                          & 0.3146                                                                         \\
RS-AvastConv                       & 98.39                           & 1.4075                                                                          & 0.0309                                                                         & 98.39                           & 7.3287                                                                          & 0.2312                                                                         \\ \hline
RsDel-MalConv                      & 98.17                           & \textbf{0.2736}                                                & \textbf{0.0275}                                                                         & 98.28                           & \textbf{1.4236}                                                & \textbf{0.1473}                                                                         \\
RsDel-MalConvGCT                   & 98.99 & 1.3270                                                                          & 0.0408                                                                         & 99.05 & 7.3082                                                                          & 0.3247                                                                         \\
RsDel-AvastConv                    & 98.51                           & 1.2751                                                                          & 0.0299                                                                         & 98.60                           & 5.4106                                                                          & 0.2205                                                                         \\ \hline
\end{tabular}%
}
\end{table}

\end{document}